\documentclass[sigplan,10pt]{acmart}
\usepackage{xspace}
\usepackage{mathtools} % upgrade from amsmath
\usepackage{bm}
\usepackage{enumitem}
\usepackage{textcomp}
\usepackage{xcolor}
\usepackage{changepage}
\usepackage{diagbox}

\usepackage{graphicx}
\usepackage{subcaption}
\usepackage{tikz}
\usepackage{caption}

\usepackage{tabularx}
\usepackage{multirow}
\usepackage{booktabs}
\usepackage{makecell}
\usepackage[figuresleft]{rotating}
\usepackage[flushleft]{threeparttable}

\usepackage{algorithmic}
\usepackage[ruled,vlined,linesnumbered]{algorithm2e}

\usepackage[utf8]{inputenc} % Avoid insert Chinese characters
\usepackage{tcolorbox} % Colored and framed text boxes
\usepackage{soul} % Colored and framed text 
\usepackage{pifont} % inconsolata
\usepackage{bbding} % icons
\usepackage[textsize=footnotesize, linecolor=magenta, bordercolor=magenta, backgroundcolor=white, textwidth=35pt]{todonotes}
\usepackage{marginnote}
\renewcommand{\marginpar}{\marginnote} %To support add note in table
\usepackage[noindentafter]{titlesec}
\titlespacing\section{0pt}{5pt}{3pt} %{left}{before}{after}
\titlespacing\subsection{0pt}{3pt}{3pt}
\titlespacing\subsubsection{0pt}{3pt}{3pt}

\usepackage[skip=3pt]{caption} 
\usepackage{ulem}
\usepackage{tikz} 
\usepackage{float}
\newcommand{\blackcircletext}[1]{%
  \tikz[baseline=(char.base)]{
\node[shape=circle,fill=black,text=white,inner sep=2pt] (char) {#1};
  }
}

\renewcommand\footnotetextcopyrightpermission[1]{}

\newcommand{\Sys}{\textit{Zeppelin}\xspace}

\AtBeginDocument{%
  }

\setcopyright{acmlicensed}
\acmYear{2026}\copyrightyear{2026}
\acmConference[EUROSYS '26]{European Conference on Computer Systems}{April 27--30, 2026}{Edinburgh, Scotland Uk}
\acmBooktitle{European Conference on Computer Systems (EUROSYS '26), April 27--30, 2026, Edinburgh, Scotland Uk}
\acmDOI{10.1145/3767295.3769369}
\acmISBN{979-8-4007-2212-7/26/04}

\acmSubmissionID{927}

\SetKwRepeat{Do}{do}{while}
\begin{document}
\sloppy
%Paint It, Black: 
\title{Zeppelin: Balancing Variable-length Workloads in Data Parallel Large Model Training}


\author{Chang Chen}
\email{charlie_chen@pku.edu.cn}
\orcid{0009-0008-6416-7172}
\affiliation{%
  \institution{Peking University}
  \country{ }
}

\author{Tiancheng Chen}
\email{tiancheng.chen@inf.ethz.ch}
\orcid{0009-0002-8071-2552}
\affiliation{%
  \institution{ETH Zurich}
  \country{ }
}

\author{Jiangfei Duan}
\email{dj021@ie.cuhk.edu.hk}
\orcid{0000-0002-6327-2033}
\affiliation{%
  \institution{The Chinese University of Hong Kong}
  \country{ }
}

\author{Qianchao Zhu}
\email{dysania@pku.edu.cn}
\orcid{0009-0001-5021-2912}
\affiliation{%
  \institution{Peking University} 
  \country{ }
}

\author{Zerui Wang}
\email{wangzerui@pjlab.org.cn}
\orcid{0009-0003-8081-8469}
\affiliation{%
  \institution{Shanghai AI Laboratory}
  \country{ }
}

\author{Qinghao Hu}
\email{qinghao.hu@ntu.edu.sg}
\orcid{0000-0003-1034-7502}
\affiliation{%
  \institution{Nanyang Technological University}
  \country{ }
}

\author{Peng Sun}
\email{sunpeng@pjlab.org.cn}
\orcid{0000-0001-8456-0491}
\affiliation{%
  \institution{Shanghai AI Laboratory}
  \country{ }
}

\author{Xiuhong Li}
\email{lixiuhong@pku.edu.cn}
\orcid{0000-0002-4896-121X}
\affiliation{%
  \institution{Peking University}
  \country{ }
}

\author{Chao Yang}
\email{chao_yang@pku.edu.cn}
\orcid{0000-0001-7426-6248}
\affiliation{%
  \institution{Peking University}
  \country{ }
}

\author{Torsten Hoefler}
\email{htor@ethz.ch}
\orcid{0000-0002-1333-9797}
\affiliation{%
  \institution{ETH Zurich}
  \country{ }
}


\renewcommand{\shortauthors}{Chang et al.}

\begin{abstract}
Training large language models (LLMs) with increasingly long and varying sequence lengths introduces severe load imbalance challenges in large-scale data-parallel training. Recent frameworks attempt to mitigate these issues through data reorganization or hybrid parallel strategies. However, they often overlook how computational and communication costs scale with sequence length, resulting in suboptimal performance.
We identify three critical challenges: (1) varying computation-to-communication ratios across sequences of different lengths in distributed attention, (2) mismatch between static NIC–GPU affinity and dynamic parallel workloads, and (3) distinct optimal partitioning strategies required for quadratic attention versus linear components.

\noindent To address these challenges, we present \Sys, a novel training system that integrates three key techniques: (1) a hierarchical sequence partitioning method for the attention module that reduces communication overhead and balances computation, supported by an efficient attention engine that applies divergent parallel strategies; (2) a routing layer that orchestrates inter-node transfers to fully utilize NIC bandwidth; and (3) a remapping layer that transforms sequence layouts between attention and linear modules, ensuring high computational efficiency across both.
Comprehensive evaluations across diverse configurations show that \Sys delivers an average 2.80× speedup over state-of-the-art methods.
\end{abstract}


\begin{CCSXML}
<ccs2012>
   <concept>
       <concept_id>10010147.10010178.10010219</concept_id>
       <concept_desc>Computing methodologies~Distributed artificial intelligence</concept_desc>
       <concept_significance>300</concept_significance>
       </concept>
 </ccs2012>
\end{CCSXML}




\maketitle

\section{Introduction}
\begin{figure}[H]
  \includegraphics[width=\linewidth]{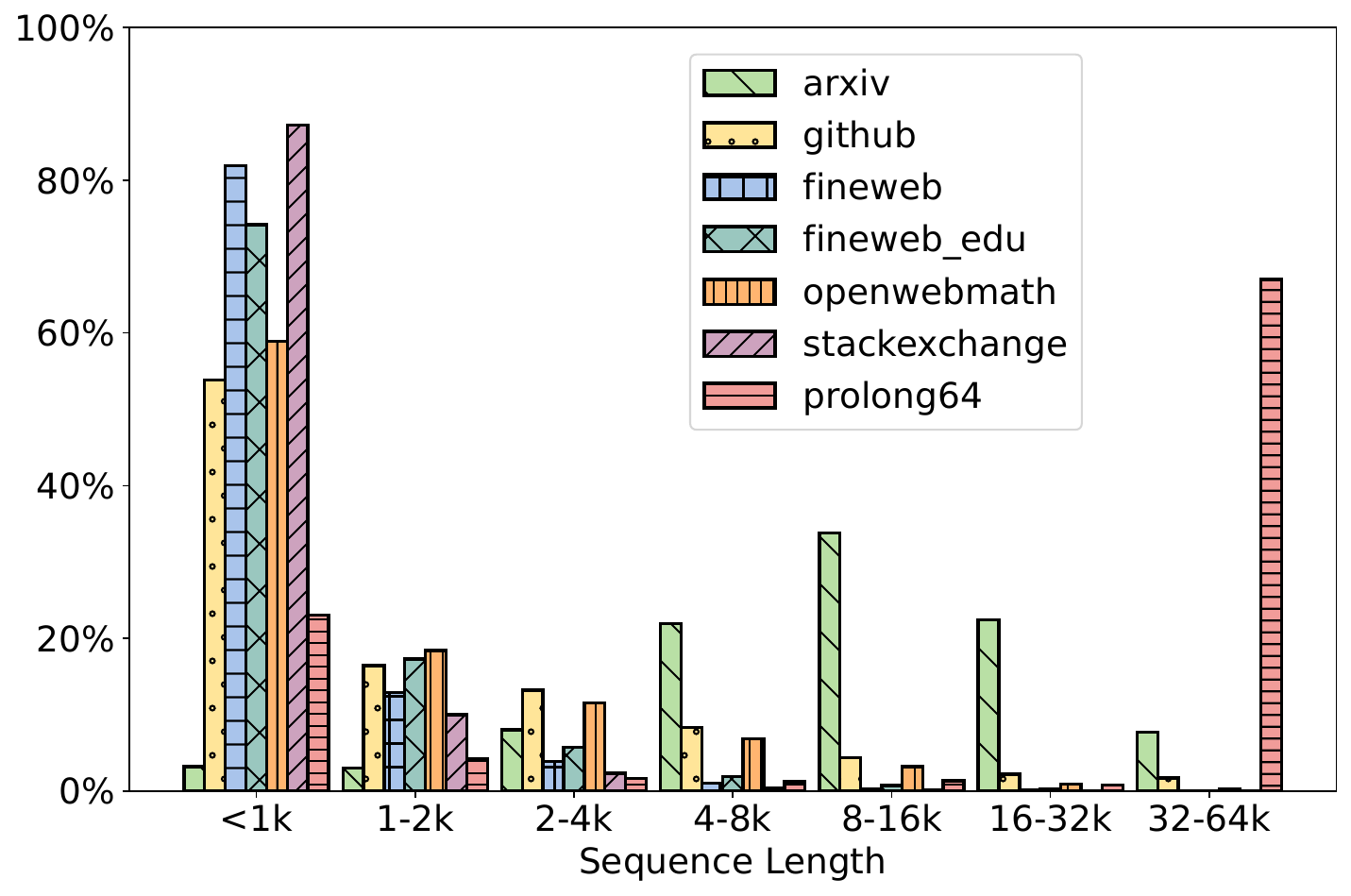}
  \caption{Sequence length distribution in multiple datasets: typical LLM training involves a mixture of datasets with diverse and often long-tailed sequence length distributions~\cite{penedo2024finewebdatasetsdecantingweb, paster2023openwebmathopendatasethighquality, kocetkov2022stack3tbpermissively, shen2024slimpajamadcunderstandingdatacombinations, 64kshormixture}.}
  \label{fig:dataset}
\end{figure}

\begin{figure*}[t]
  \includegraphics[width=\textwidth]{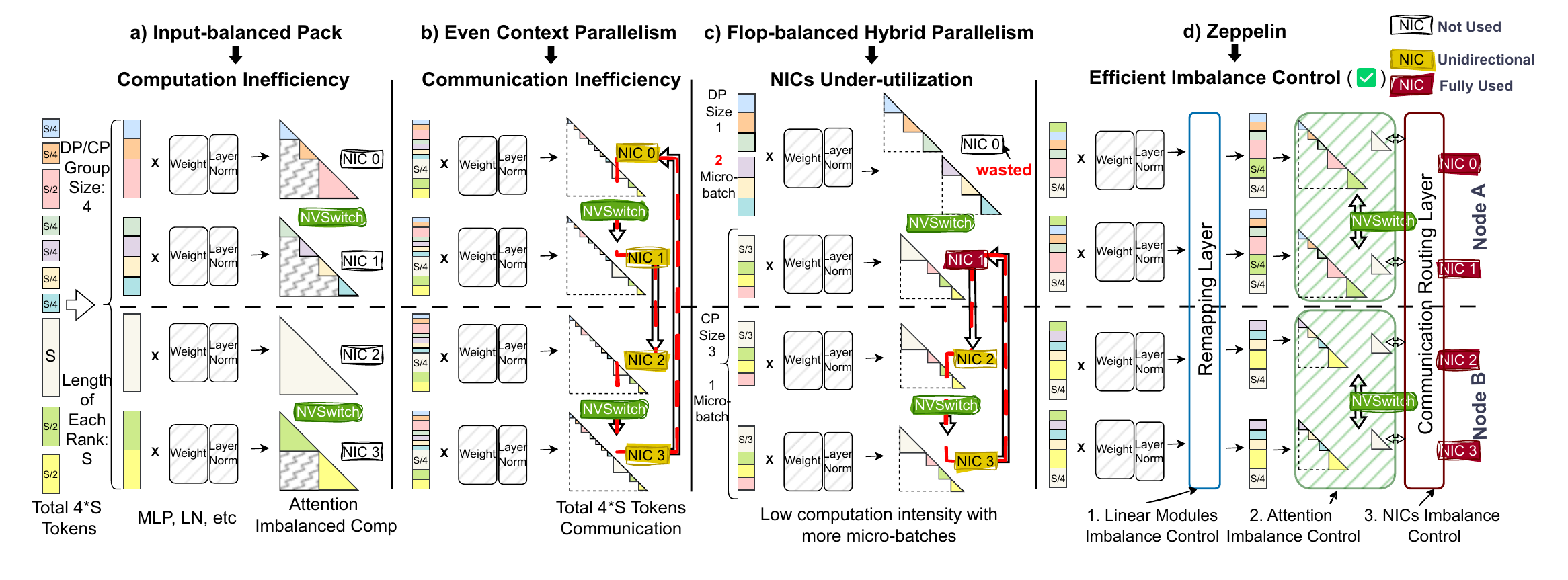}
  \caption{Balancing a complex system often propagates inefficiency internally.
a) Input balancing \cite{deepseekai2025deepseekv3technicalreport, qwen2025qwen25technicalreport} struggles with quadratic scaling of attention, leading to computation inefficiency.
b) Even sequence splitting \cite{te, grattafiori2024llama3herdmodels, wlb} balances computation but incurs high communication overhead, especially for short sequences.
c) Hybrid parallel methods \cite{flexsp, bytescale} lead to uneven hardware (NIC) utilization and computational intensity across ranks.
d) \Sys addresses imbalance holistically from the model to hardware.}
  \label{fig:intro}
\end{figure*}
Data parallelism (DP) is a well-established paradigm for training large language models (LLMs). It distributes large mixtures of datasets across multiple devices, with each device processing its local batch using an identical model replica. 
%% variable length, long context challenges
Two recent trends in pretraining data mixture~\cite{grattafiori2024llama3herdmodels,qwen2025qwen25technicalreport,deepseekai2025deepseekv3technicalreport, gemini} are the use of much longer sequences and the necessity of training on mixtures of datasets with highly variable sequence lengths, as shown in Fig.~\ref{fig:dataset}. 
Longer sequences are crucial for capturing complex dependencies and improving model capabilities~\cite{grattafiori2024llama3herdmodels, gemini, bertsch2023unlimiformerlong}, but they impose a quadratic computational cost in the self-attention module. Variable-length sequences are essential for models to handle diverse inputs, from short queries to long documents~\cite{mixture1fuyao, mixture2, 64kshormixture, mixture3, llama2_mixtrue}. Together, these long and variable-length inputs introduce severe load imbalance across DP ranks when the context window scales to hundreds of thousands of tokens, as seen in recent large-scale distributed training recipes~\cite{grattafiori2024llama3herdmodels,qwen2025qwen25technicalreport,deepseekai2025deepseekv3technicalreport,bytescale, wlb}.

As illustrated in Fig~\ref{fig:intro}, recent methods attempt to mitigate load imbalance through data reorganization or hybrid data parallelism, typically optimizing for a balance metric.
\textbf{Input-balanced pack} (Fig.~\ref{fig:intro}.a) creates uniformly sized input tensors at the start of each iteration using techniques such as sequence packing or chunking~\cite{deepmindpack, ding2024fewertruncationspackingimprovelanguage, applebucket, zhao2024chunk}. This approach is adopted in the Qwen~\cite{yang2024qwen2technicalreport, qwen2025qwen25technicalreport} and DeepSeek~\cite{deepseekai2024deepseekv2strongeconomicalefficient, deepseekai2025deepseekv3technicalreport} model families. While effective for balancing computation in linear modules, it often lead to redundant attention computation or imbalanced attention masking.
\textbf{Even context parallelism (CP)} (Fig.~\ref{fig:intro}.b) splits each sequence across devices to balance attention computation, employing distributed attention mechanisms method~\cite{ulysses,megatronsp,gu2024loongtrainefficienttraininglongsequence,liu2023ringattentionblockwisetransformers, grattafiori2024llama3herdmodels, brandon2023stripedringattentionfasterring}. This strategy is adopted in LLaMA 3~\cite{grattafiori2024llama3herdmodels, wlb} training and the Megatron-LM framework \cite{megatron,te, chen2025longvila}. However, it introduces substantial communication overhead, particularly for the numerous short sequences common in diverse datasets.
To reduce communication costs, \textbf{flop-balanced hybrid parallelism} (Fig.~\ref{fig:intro}.c) combines DP and CP, assigning short and long sequences to separate micro-batches~\cite{flexsp, bytescale,hotswitch}. While this balances end-to-end flop across devices, DP ranks become NIC-underutilized when processing more micro-batches of short sequences, whereas CP ranks remain computation-heavy but memory-underutilized.

%% 3 scaling gap analysis
Despite aiming to balance computation among devices, existing approaches often overlook the distinct scaling behaviors of LLM modules with respect to sequence length. 
First, different modules exhibit different complexities: linear moduules (e.g., MLP) scale linearly, while attention scales quadratically with sequence length~\cite{attention,gpt}. An input distribution that is balanced for linear modules becomes inefficient for quadratic ones when length distribution is highly dynamic. Second, distributed attention incurs communication volume proportional to sequence length~\cite{ulysses, li2024distflashattndistributedmemoryefficientattention, 
grattafiori2024llama3herdmodels}, leading to large discrepancies in computation-to-communication ratios across variable-length sequences. This indicates that such sequences place different demands on both computational resources and heterogeneous interconnects. However, in modern training systems each GPU is connected to a NIC \cite{a100,h100} through a PCIe switch, 
and these network resources are often underutilized in distributed attention.
Given these insights, the performance of a training system can be significantly compromised by variable-length inputs, and optimizations based solely on a balance metric cannot deliver optimal performance.

\begin{figure*}[t]
  \includegraphics[width=\textwidth]{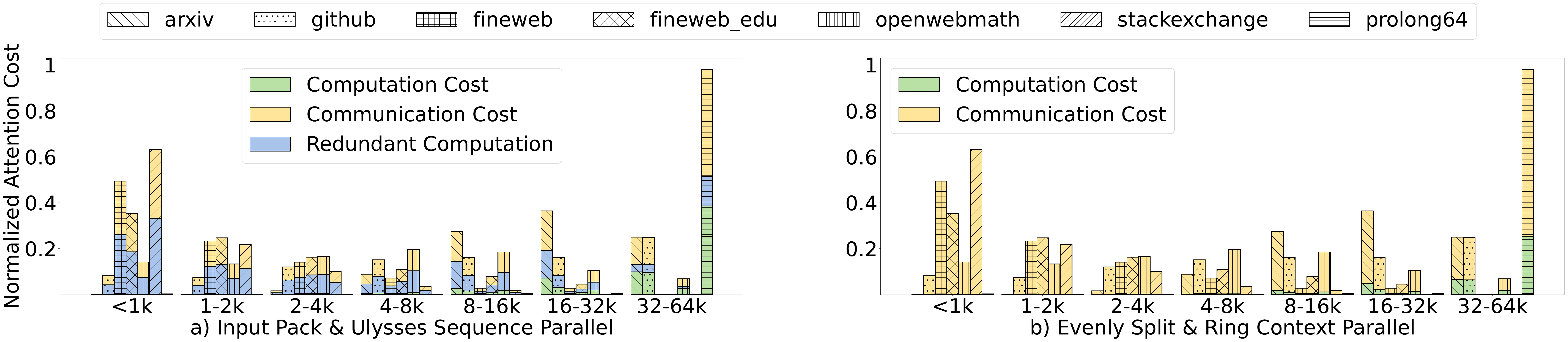}
  \caption{Multi-head attention cost distribution across different sequence length ranges in each dataset (normalized to the total attention cost per dataset). Evaluation on a 2-node system with 8 A800 GPUs per node, a total sequence length of 64k, and 4$\times$200 Gbps NICs per node. (a) Packing-based input balancing introduces significant redundant computation, especially for short sequences.
b) CP with even sequence splitting incurs substantial unnecessary communication for short sequences, and cross-node communication is difficult to fully overlap with computation.}
  \label{fig:motivation}
\end{figure*}

%Analisys

We present \Sys, a balanced and efficient data-parallel LLMs training system for variable-length sequences. \Sys incorporates three key design innovations:

First, mismatched scaling trends between computation and communication in distributed attention necessitate a fine-grained hybrid parallel strategy. To leverage the hierarchical bandwidth structure of modern interconnects, we categorize context parallel strategies into three types: local, intra-node, and inter-node, and propose a two-step, topology-aware sequence partitioning and placement method. Building on this partitioning, a flexible attention engine orchestrates the execution of these sequence types, ensuring efficient overlap between computation and communication.

Second, the dynamic communication patterns of variable-length sequences require disaggregating the fixed affinity between GPUs and NICs within each node. To fully utilize all NICs, we propose a three-step routing mechanism that substitutes direct inter-node communication with (1) intra-node dispatch, (2) multi-NIC inter-node transfer, and (3) intra-node combine. This design alleviates inter-node bandwidth bottlenecks and enables overlap of computation with both intra- and inter-node data transfers.

Third, balancing computation in linear modules requires a remapping mechanism aligned with the attention partitioning scheme. The remapping layer adjusts sequence layouts before and after linear modules, ensuring balanced workloads. We formalize this as an optimization problem to minimize remapping communication cost.

% design
To conclude, this paper makes the following contributions:
\begin{itemize}[leftmargin=*] 
\item \textbf{Characterization of scaling behavior.} We analyze how key components scale with sequence length, including quadratic modules (attention), linear modules (MLPs), distributed attention with linear communication cost, and hardware-level GPU-NIC affinity.
\item \textbf{System design.} We propose a bandwidth-aware partitioning strategy that categorize sequences into local, intra-node, and inter-node types. Based on this, We design and implement a distributed attention engine to efficiently manage parallel execution. To accelerate dynamic communication workloads, we introduce a three-step routing scheme that fully utilizes all NICs within a node, overlapping computation with intra- and inter-node communication. Additionally, we design a remapping layer that balances workloads in linear modules with minimal overhead. 
\item \textbf{Comprehensive evaluation.} We conduct extensive experiments across diverse models, sequence lengths, and training scales. Results demonstrate that \Sys achieves an average speedup of \textbf{2.80}$\times$ over state-of-the-art methods.
\end{itemize}

%% contribution

\section{Background and Motivation}

\subsection{Transformer Architecture}
LLMs based on the transformer architecture typically compromise a stack of transformer layers \cite{qwen2025qwen25technicalreport, grattafiori2024llama3herdmodels, gpt}, each containing a quadratic self-attention module alongside several linear computation modules. The self-attention mechanism captures contextual dependencies across the entire sequence, introducing quadratic computational complexity with respect to sequence length \cite{attention}. Owing to causal dependencies among tokens, the attention mask typically follows a lower-triangular pattern. In contrast, operations such as normalization, linear projection, and activation functions are token-wise, enabling each token to be processed independently. These operations are referred to as "linear modules" collectively in this paper.

\subsection{Data Parallel Variants}
Traditional data parallel methods pack input tensors along the batch dimension. Before the era of large models, data parallelism primarily focused on increasing batch size to improve throughput~\cite{lars,lamb, goyal2018accuratelargeminibatchsgd, largebatch}. When inputs were small, simple padding could easily achieve load balance without harming training performance ~\cite{padding, huggingface}.

\begin{figure*}[t]
  \includegraphics[width=\textwidth]{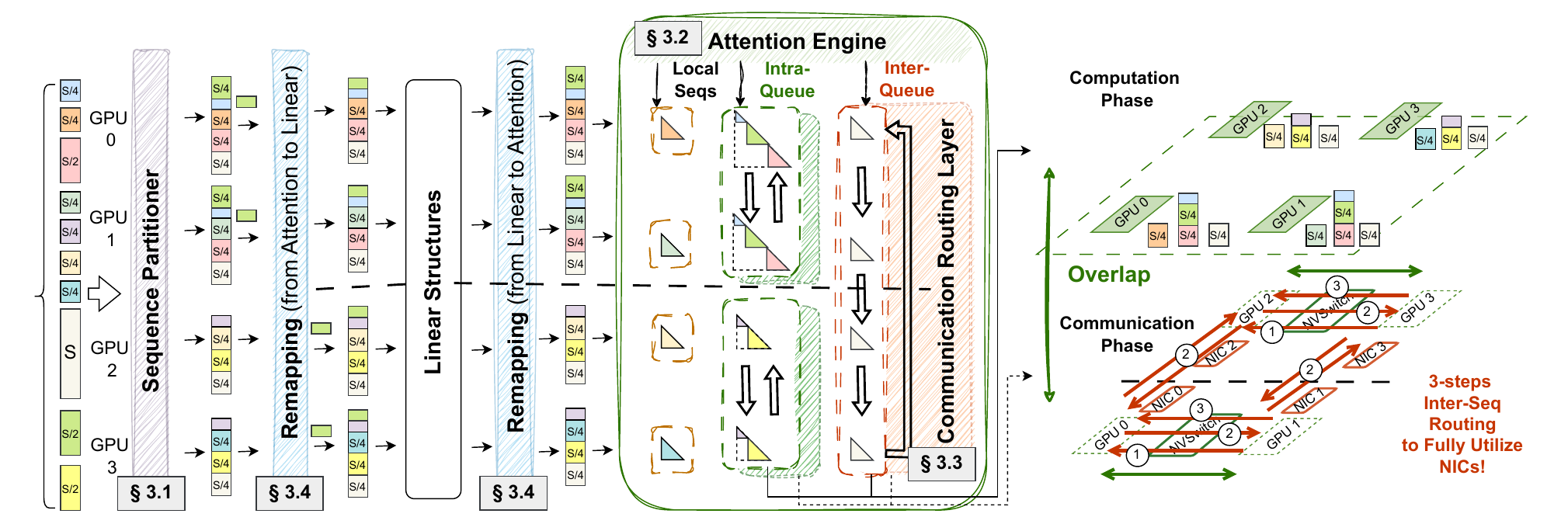}
  \caption{Overview of \Sys: The Sequence Partitioner assigns input sequences to minimize imbalance caused by quadratic attention. Before entering linear modules, sequences are remapped into a linearly balanced pattern and then restored after linear computation. The Attention Engine orchestrates three types of sequence queues, overlapping computation with communication. Meanwhile, the Communication Routing Layer optimizes inter-node data transfers by replacing direct cross-node communication with a three-step routing mechanism, ensuring full utilization of available NIC bandwidth.}
  \label{fig:main}
\end{figure*}

In transformer-based large model training, however, both sequence length and model size have increased significantly. To meet these demands, various distributed training strategies have been developed, broadly classified into two orthogonal types: Model Parallelism (splitting model weights, e.g., tensor and pipeline parallelism~\cite{megatron,megatron2}) and Data Parallelism variants. Key DP variants include ZeRO methods~\cite{deepspeed}, Sequence Parallelism (SP)~\cite{megatronsp}, and Context Parallelism (CP)~\cite{te}. Megatron-LM's SP~\cite{megatronsp} inserts all-gather and reduce-scatter operations to synchronize intermediate activations, while Ulysses SP~\cite{ulysses} in DeepSpeed~\cite{deepspeed} employs all-to-all communication to better manage communication and memory, though it requires the group size to be a factor of the number of attention heads. CP in Transformer Engine~\cite{te} utilizes ring attention, processing queries iteratively with different key-value chunks while overlapping send-receive operations via a ring topology~\cite{liu2023ringattentionblockwisetransformers, li2024distflashattndistributedmemoryefficientattention}. Many recent DP variants are flexible combinations of these base methods~\cite{gu2024loongtrainefficienttraininglongsequence,usp}.

These sequence-parallel approaches can be further adapted to dynamic-length input. Packing and padding strategies naturally integrate with SP, while evenly splitting each sequence aligns well with CP~\cite{te, chen2025longvila}. Hybrid DP methods assign sequences of different lengths to appropriate variants and balance floating-point operations (flop) across ranks to reduce communication overheads~\cite{flexsp, bytescale,wlb}.

\subsection{Load Balance, at What Cost?}
 %\todo{Better to have a quantitative comparison of different strategies for Figure 2 here}
All DP variants discussed above aim to balance training with respect to a chosen metric, such as memory usage or computational flop. However, these "balanced" approaches inevitably introduce new source of overhead, as illustrated in Fig.~\ref{fig:motivation}. We identify three major inefficiencies underlying these methods:
\begin{itemize}[leftmargin=*] 
    \item \textbf{Computation Inefficiency}: Input-balanced packing approaches successfully balance linear computations but introduce significant inefficiency in attention. As shown in Fig.~\ref{fig:motivation}.a, packing wastes a substantial portion of computation on cross-sequence attention and incurs high communication cost. In datasets with many short sequences, redundant cross-sequence computation and communication dominate attention overhead, reaching up to 60\% for sequences shorter than 1k tokens in StackExchange~\cite{kocetkov2022stack3tbpermissively}, as shown in Fig.~\ref{fig:motivation}.a.
    
    \item \textbf{Communication Inefficiency}: Evenly partitioning sequences across devices balances both memory and computation but incurs excessive communication overhead, especially for batches with many short sequences. As shown in Fig.~\ref{fig:motivation}.b, the low computation-to-communication ratio of short sequences severely limits performance in distributed attention. 
    Consequently, a single parallel strategy cannot effectively handle batches with diverse sequence lengths under heterogeneous inter- and intra-node bandwidth conditions.

    \item \textbf{Hardware Under-utilization}: Hybrid strategies that combine multiple DP variants attempt to reduce communication costs by splitting batches across different DP groups. However, this introduces imbalanced hardware utilization and computation intensity across parallelism strategies. As illustrated in Fig.~\ref{fig:intro}.c, this often requires more micro-batches to balance memory and flop, reducing per-microbatch token count and overall compute intensity. Moreover, mixing different DP schemes frequently leads to uneven utilization of NICs. Achieving optimal performance thus requires simultaneously maintaining high compute intensity and high NIC utilization.
    % Strategies that combine multiple DP variants can lead to inefficient hardware utilization (Fig.~\ref{fig:intro}.c). These strategies typically split a data batch and assign it to different DP groups to reduce communication costs. However, they often require an increased number of micro-batches to balance memory usage and flop, which reduces the number of tokens processed per micro-batch and leads to lower computation intensity. Additionally, combining different DP variants can result in imbalanced utilization of NIC resources. Therefore, maintaining high computation intensity and ensuring efficient, balanced utilization of all NICs are critical to achieving optimal overall system performance.

\end{itemize}

% \section{Libra Design}
\section{Zeppelin Design}

Building on the above analysis, we argue that achieving comprehensive and efficient load balance with respect to a single metric is inherently difficult. Instead, we identify structural inefficiencies in existing data-parallel training system and propose a new system design, \Sys, to address them holistically.
As illustrated in Fig.~\ref{fig:main}, \Sys incorporates four key design components that collectively mitigate load imbalance across the training system:

% \noindent $\textcircled{1}$ 
\noindent \blackcircletext{1}  \textbf{Sequence Partitioner} (\S~\ref{subsec:seq_par}) splits and places sequences using a two-level hierarchical strategy, accounting for the differing computation and communication demands of variable-length sequences in distributed attention.  

\noindent \blackcircletext{2}  \textbf{Attention Engine} (\S~\ref{subsec:attn_eng}) 
dynamically manages sequence execution across local, intra-node, and inter-node tiers, adapting to the diverse execution characteristics of different sequence parallel types.

\noindent \blackcircletext{3}  \textbf{Communication Routing Layer} (\S~\ref{sec:hardware}) addresses imbalances in communication patterns by decoupling GPU-NIC affinity. This enables better utilization of communication resources and improved overlap of computation and communication.

\noindent \blackcircletext{4}  \textbf{Remapping Layer} (\S~\ref{remapper}) improves performance by dynamically adjusting placement of sequences for linear structures with small data transfer overheads, guided by the attention partitioning pattern.

%The remainder of this section is organized as follows: first, an optimization problem formulation for variable-length sequence training; second, partition, placement, and remapping strategies; and third, detailed designs of the attention runtime and communication routing layer.
% Sequence Partitioner
% Remapper
% Attention Runtime Manager
% Communication Router 

\begin{table}
  \caption{Notations}
  \label{tab:nota}
  \begin{tabular}{cl}
    \toprule
    
    $S$ &  Total Number of Sequences\\
    $s_i$ &  $i$-th Sequence Length\\
    $P$ &  Number of Devices per Node \\
    $N$ &  Number of Nodes \\
    $L$ &  Token Capacity of Each GPU \\
    %$\alpha$ & attention computation cost coefficient\\
    %$\beta$ & linear computation cost coefficient\\
    $b_{\text{inter}},b_{\text{intra}}$ & inter-, intra- inverse bandwidth cost \\
  \bottomrule
\end{tabular}
\end{table}

\subsection{Sequence Partitioner}
\label{subsec:seq_par}

\begin{figure}
  \includegraphics[width=\linewidth]{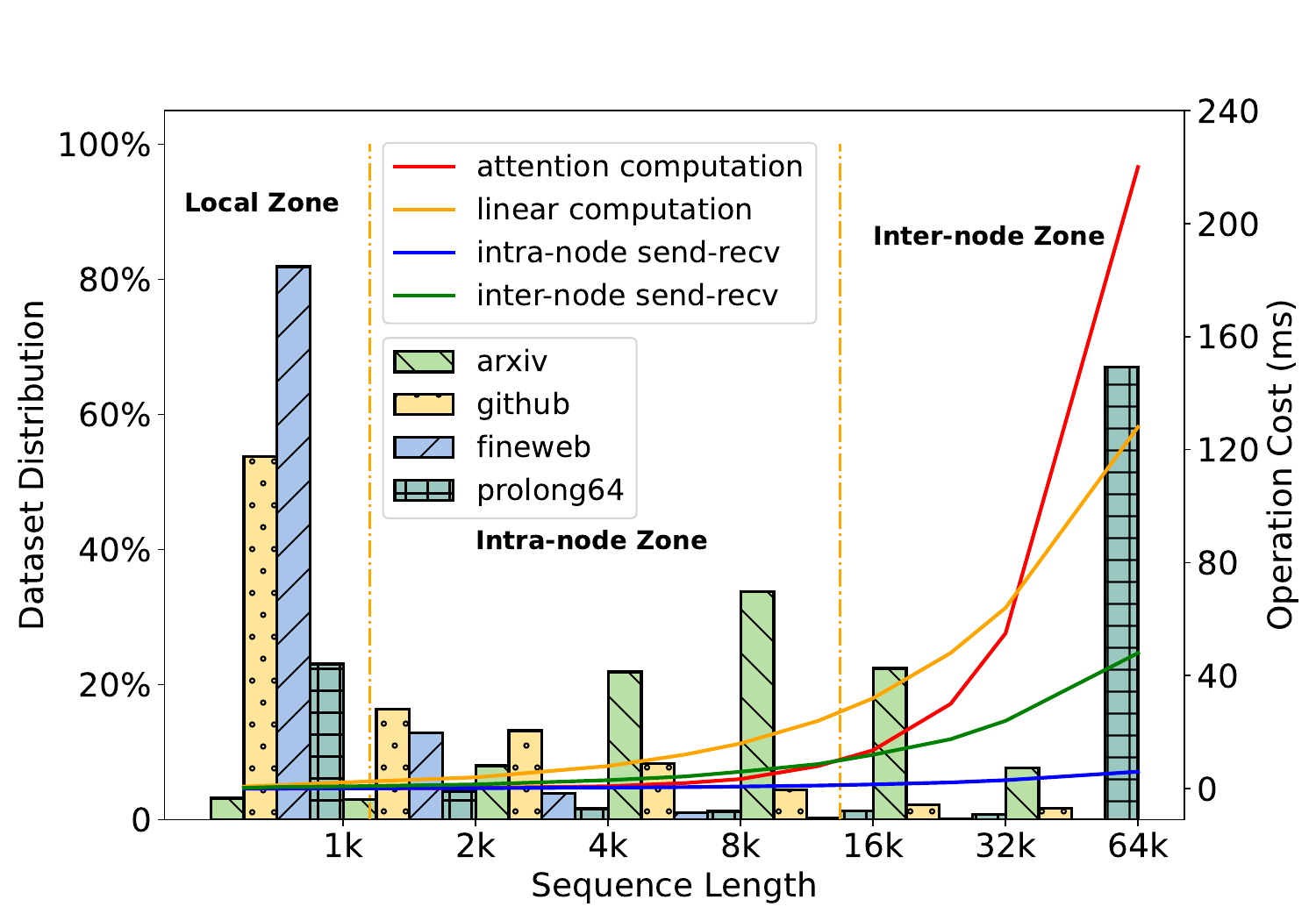}
  \caption{Attention computation cost on an A800 GPU and send-receive cost under 200 Gb/s inter-node and 400 GB/s intra-node bandwidths for different sequence lengths. The intersections of the three cost curves divide datasets into three zones: local, intra-node, and inter-node, representing the overlapping capabilities of sequences of different lengths.}
  \label{fig:3zones}
\end{figure}

The end-to-end training cost compromises two main components: attention modules and linear modules. As discussed earlier, efficient sequence splitting and placement strategies differ between these components, particularly under dynamic input length distributions. Since quadratic attention constitutes the primary bottleneck in long-sequence training (Fig.~\ref{fig:3zones}), we begin our analysis with the attention module. The analysis of linear modules, together with the associated remapping strategy, is deferred to Section ~\ref{remapper}. 

In practice, quadratic attention computation is distributed across a group of ranks to alleviate computation overhead. However, this distributed attention pattern requires transferring large volumes of key-value (KV) activations. The direct Allgather approach, adopted by LLaMA~3 training~\cite{grattafiori2024llama3herdmodels, wlb}, increases peak memory usage and places communication on the critical path. To improve overlap and reduce memory pressure, communication can instead be decomposed into rounds of send-receive operations, which interleave naturally with sharded attention computation. The send-receive-based ring attention mechanism~\cite{liu2023ringattentionblockwisetransformers} provides greater flexibility, offering a more adaptable foundation for per-sequence parallelism by supporting dynamic group sizes and fine-grained partitioning, compared to a single-synchronization Allgather across a rank group. 

%To simplify the solving process and derive 

Based on the ring attention pattern,
we identify a core objective in sequence splitting and placement: maximizing overlap between computation and communication over a hierarchical network. As shown in Fig.~\ref{fig:3zones}, computation scales more rapidly than communication, yielding a computation-to-communication ratio that increases linearly with sequence length. This trend indicates that longer sequences can more effectively hide communication costs, while  shorter sequences suffer from inefficient communication. Leveraging this insight, we categorize sequences into three distinct zones, as depicted in Fig.~\ref{fig:3zones}.
\begin{itemize}[leftmargin=*]
    \item \textbf{Local zone:} Short sequences that require no communication  and are most efficiently processed on a single device. 
    \item \textbf{Intra-node zone:} Medium-length sequences that benefit from overlapping intra-node communication with computation.
    \item \textbf{Inter-node zone:} Long sequences that span multiple nodes, where high computation cost effectively overlaps inter-node communication.
\end{itemize}
This three-zone categorization aligns with the bandwidth hierarchy of modern GPU training infrastructures, where inter-node bandwidth is typically an order of magnitude lower than intra-node bandwidth. 
With this model, the optimization problem reduces to two key decisions: $\textcircled{1}$ determining the appropriate zone for each sequence and $\textcircled{2}$ partitioning and placing sequences within each zone at the proper granularity. 

\begin{algorithm}[tb]
    \SetAlgoLined
    \SetKwInOut{Input}{Input}
    \SetKwRepeat{Do}{do}{while}
    \Input{ Input Sequences $\mathcal{S}$,
    Devices per node $P$,
    Token capacity per device $L$, Number of Nodes $N$,}
    %s = Sum($\{s_1^2, s_2^2,..., s_S^2\}$) / $N$ \; 
    Sort $\mathcal{S}$ in descending order by sequence length\;
    Initialize inter-node zone threshold $s_1 = P*L$\;
    %Separate $\mathcal{S}$ into $\mathcal{S}_1, \mathcal{S}_2$ according to $S_2$\; 
      
    \Do{
        flag
    }{
    %/* \textcolor{blue}{Inter-node sequences split} */ \\
            $node\_buckets$ = [[] x $N$], flag = \texttt{False}  \;
            intra-node/local zone $z_{01} = \{|\mathbf{s}| < s_1|\mathbf{s} \in \mathcal{S}$\}\;
            inter-node zone $z_2 = \{|\mathbf{s}| \geq s_1|\mathbf{s} \in \mathcal{S}\}$ \;
            per-node length budget $s_{\text{avg}}=\sum_{\mathbf{s} \in z_2}{|\mathbf{s}|}/N$ \;
            \For{$\mathbf{s}$ in $z_2$}{
                % evenly split $s$ into $|s|/s_{avg}$ empty buckets\;
                Split $\mathbf{s}$ evenly to $\left\lceil |\mathbf{s}| / s_{\text{avg}} \right\rceil$ empty buckets\;
            %Distribute fragments evenly across buckets\;
            }
        
    %/* \textcolor{blue}{Intra-node  and local sequences placement} */ \\
            \For{$\mathbf{s}$ in $z_{01}$}{
               $idx$ = $argmin_{i}(\sum_{ss \in node\_buckets[i]}|ss|$)\;
               \eIf{$|\mathbf{s}|+ \sum_{ss \in node\_buckets[idx]}{|ss|} > P*L$}{
                Update threshold: $s_1 = max\{z_{01}\}$\;
                flag = \texttt{True}; \textbf{break}\;
               }{
               $node\_buckets[idx]$.push($\mathbf{s}$)\;
               }
            }
    }
    \SetKwInOut{Output}{Output}
    \Output{$node\_buckets$}
    \caption{Inter-Node Partitioning\label{alg:1}}
\end{algorithm}

We propose a hierarchical partitioning strategy consisting of two steps: inter-node partitioning followed by intra-node partitioning. As illustrated in Alg.~\ref{alg:1}, the inter-node stage determines the boundary $s_1$ between the inter-node $z_2$ and intra-node/local $z_{01}$ zones, and assigns them into $N$ node-level buckets to optimize communication, which is the primary bottleneck at this level.
The algorithm iteratively adjusts $s_1$: it is initially set by the per-node token capacity (Line 2), and inter-node sequences $z_2$ are then chunked based on the average per-node budget $s_{\text{avg}}$ to balance load. Instead of spreading sequences uniformly across all nodes, we increase the partition granularity for cross-node sequences and assign them to separate GPUs (Lines 7–10), improving communication efficiency. The remaining shorter sequences in $z_{01}$ are assigned to the least-loaded node buckets (Lines 11–15). If any sequence in $z_{01}$ exceeds node capacity (Lines 13–15), $s_1$ is reduced to the maximum of $z_{01}$ and the process repeats. The iterative refinement guarantees that all sequences shorter than the final $s_1$ can be placed within node capacity.

\begin{algorithm}[htb]
    \SetAlgoLined
    \SetKwInOut{Input}{Input}
    \Input{Intra-node Sequences $z_{01}$, Inter-node Sequences $z_{2}$ (at curent node), Devices per Node $P$,
    Token Capacity per Device $L$}
    %$inter\_queue$ = $inter\_queues[self\_node]$\;
    %$s1 = sum(\{len(s)^2|s \in (\mathcal{S}_1 \cap inter\_queue)\}) / P$ \;
    Initialize threshold $s_0 = L$\;
    %Separate $\mathcal{S}$ into $\mathcal{S}_1, \mathcal{S}_2$ according to $S_2$\; 
      
    \Do{
        flag
    }{
    %/* \textcolor{blue}{Inter-node sequences split} */ \\
            $device\_buckets$ = [[] x $P$], flag = \texttt{False}  \;
            \For{$\mathbf{s}$ in $z_2$ }{
            Split $\mathbf{s}$ evenly to $P$ devices\;
            }
            $z_{0} = \{|\mathbf{s}| < s_0|\mathbf{s} \in {z_{01}}\}, z_1 = \{|\mathbf{s}| \geq s_0|\mathbf{s} \in {z_{01}}\}$ \;
            per-device budget $c_{\text{avg}}=\sum_{\mathbf{s}\in z_1}(|\mathbf{s}|^2)/P$\;
            \For{$\mathbf{s}$ in $z_1$}{
                Split $\mathbf{s}$ into $\left\lceil |\mathbf{s}|^2/c_{\text{avg}} \right\rceil$ fragments\;
                Assign fragments to buckets in round-robin fashion\;
            }
        
    %/* \textcolor{blue}{Intra-node  and local sequences placement} */ \\
            \For{$\mathbf{s}$ in $z_{0}$}{
               $idx$ = $argmin_{i}(\sum_{ss \in device\_buckets[i]}|ss|$)\;
               \eIf{$|\mathbf{s}|+ \sum_{ss \in device\_buckets[idx]}|ss|) > L$}{
                Update threshold: $s_0 = max\{z_{0}\}$\;
                flag = $\texttt{True}$; $\textbf{break}$\;
               }{
               $device\_buckets[idx]$.push($\mathbf{s}$)\;
               }
            }
    }
    \SetKwInOut{Output}{Output}
    \Output{$device\_buckets$}
    \caption{Intra-Node Partitioning\label{alg:2}}
\end{algorithm}

At the intra-node stage, Alg.~\ref{alg:2} further partitions the sequences within each node across $P$ devices. Similar to Alg.~\ref{alg:1}, the algorithm determines the boundary $s_0$ between intra-node $z_1$ and local $z_0$ sequences. Inter-node sequences are evenly distributed across $P$ GPUs. Since intra-node communication overlaps more effectively with computation, intra-node sequences are split to balance their quadratic computation across devices, using $c_{\text{avg}}$ as the per-device budget. Finally, local sequences are assigned to individual devices. If any local sequence exceeds device capacity $L$, $s_0$ is adjusted iteratively to the maximum of $z_0$. This iterative adjustment guarantees feasible placement while balancing computation and communication within each node.

\subsection{Attention Engine}
\label{subsec:attn_eng}

With the sequence assignments, each device receives sequences categorized into three types: inter-node, intra-node, and local. Each type mapped to a distinct ring communication group. Compared to the evenly split pattern, where all sequences are evenly split across a single global ring~\cite{te, brandon2023stripedringattentionfasterring}, the hierarchical ring-group mapping reduces the communication overhead from $b_{\text{inter}}\sum\limits_{i\in S} s_i$ to $\max\limits_{\mathbf{s} \in z_2} b_{\text{inter}} \mathbf{s} + \max\limits_{\mathbf{s} \in z_1} b_{\text{intra}} \mathbf{s}$, where $b_{\text{inter}}$ and $b_{\text{intra}}$ denote the inverses of inter- and intra-node bandwidths, respectively. Hybrid methods~\cite{flexsp, bytescale} that partition sequences into multiple communication groups typically apply coarse-grained model-level parallelism, which often introduces imbalance among ranks. In contrast, \Sys employs a fine-grained attention-level execution engine. It orchestrates three separate queues, ensuring balanced workload distribution while aligning with the underlying network hierarchy.

\begin{figure}
  \includegraphics[width=\linewidth]{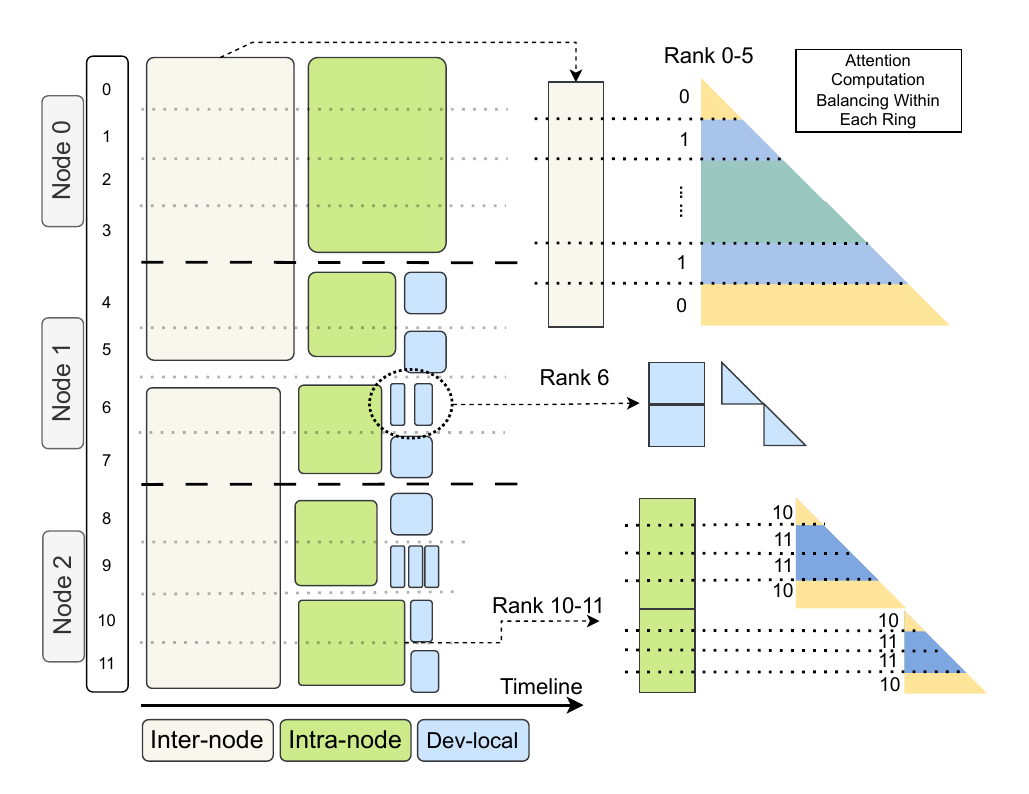}
  \caption{Attention Engine orchestrates partitioned sequences on each device, performing computations in the order of inter-node, intra-node, and local. Sequences are partitioned to balance the computational load, considering the triangular attention mask.}
  \label{fig:runtime}
\end{figure}

As shown in Fig.~\ref{fig:runtime}, execution proceeds in the order of inter-node, intra-node, and local sequences. This ordering is crucial for maximizing efficiency. 
Because the inter-node communication group spans and subsumes intra-node groups on participating nodes, completing inter-node tasks first enables immediate launching of intra-node queues without blocking. In contrast, executing intra-node tasks first would delay inter-node launches, as they must wait for all intra-node operations across nodes to finish, introducing idle time. 
Finally, local sequences, requiring no communication, are executed last to avoid interfering with communication-dependent tasks.

\begin{figure*}[t]
  \includegraphics[width=\linewidth]{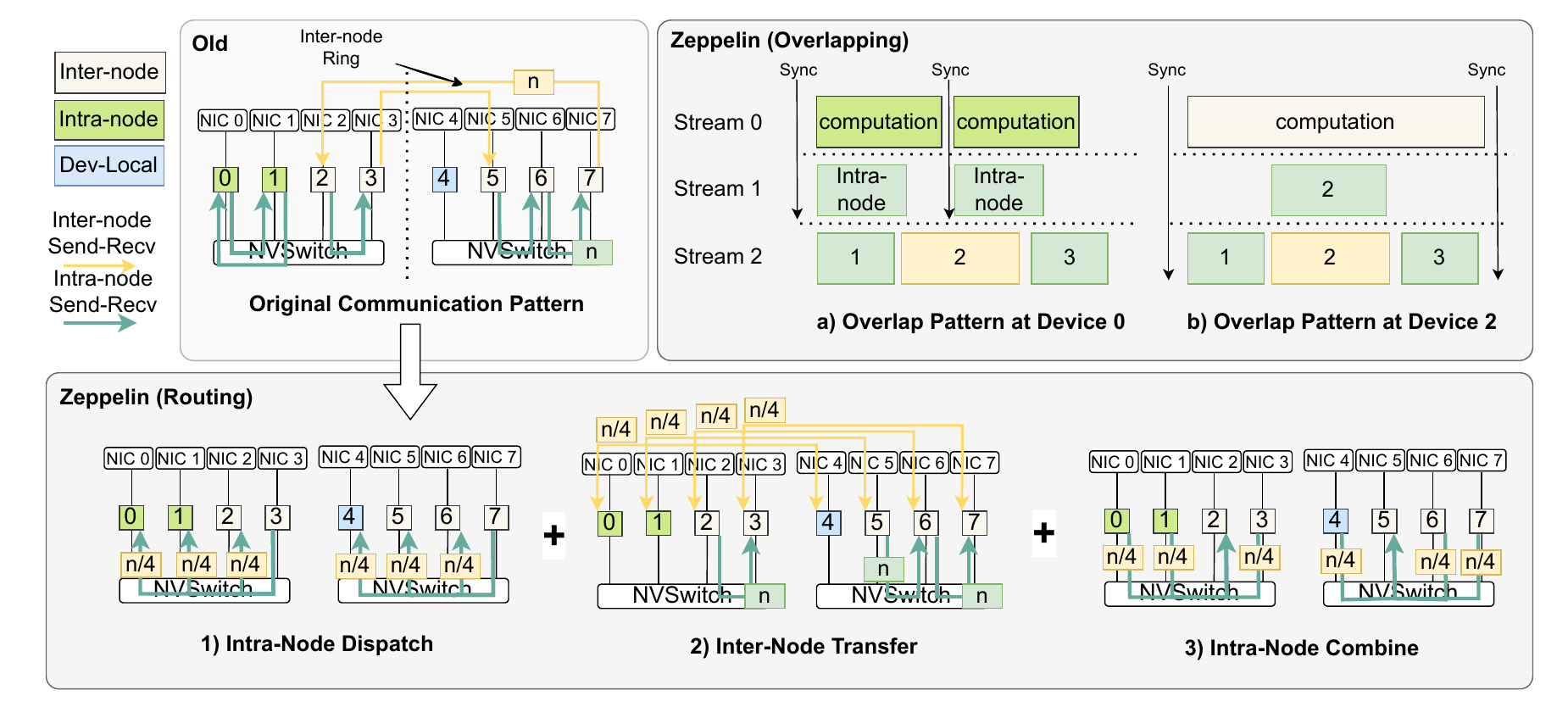}
  \caption{The send-receive operations of an inter-node ring for a data volume of $n$ are decomposed into three steps: 1) local dispatch:  the inter-node send workload is first distributed to designated proxy ranks within each node, 2) inter-node exchange: these proxy ranks perform the inter-node send-receives for the dispatched workload, overlapped with ongoing intra-node transfers, and 3) local gather: the data received from remote nodes is gathered on the target rank. Ranks assigned for intra-node and local attention computation also serve as proxy ranks to route the data.}
  \label{fig:routing}
\end{figure*}

The lower-triangular computation pattern in causal self-attention~\cite{gpt, grattafiori2024llama3herdmodels} requires a fine-grained sequence partitioning strategy for inter- and intra-node rings to achieve balanced load across participating ranks ~\cite{wlb}. As illustrated in Fig.~\ref{fig:runtime}, sequences within a queue, e.g., an inter-node ring of size $G_{\text{inter}}$ or an intra-node ring of size $G_{\text{intra}}$, are further divided into $2\times G$ equal-length chunks, where $G$ denotes the ring group size. Each rank $i$ in the ring is assigned the $i$-th and the $(2G-i-1)$-th chunks, ensuring balanced computation among all ranks in the group. Execution follows the standard ring attention pattern:   \cite{li2024distflashattndistributedmemoryefficientattention, brandon2023stripedringattentionfasterring, liu2023ringattentionblockwisetransformers}, the execution of each queue involves $G$ rounds of computation overlapped with send-receive communication. In each round, rank $i$ sends its current KV activations to rank $i+1$ and receives the next round's KV activations from rank $i-1$. %The cross-node communication is further optimized for hardware utilization as described in Sec.\ref{sec:hardware}. 
Local sequences are executed directly on their assigned GPU using a standard variable-length attention kernel without inter-rank communication.

\subsection{Communication Routing}
\label{sec:hardware}

%algoritm part for routing

The hierarchical partitioning strategy, combining local, intra-node, and inter-node parallel patterns in distributed attention, helps reduce overall communication volume. However, the flexible combination of these methods introduces diverse communication patterns that can still result in hardware under-utilization.
For example, devices processing local or intra-node sequences may leave their dedicated NICs idle while other devices on the same node engage in inter-node communication, as shown in Fig.~\ref{fig:routing}. Within an inter-node ring, the structured send-receive pattern, e.g., rank $i$ sends to $i+1$ and receives from $i-1$, can also lead to NIC under-utilization. NICs associated with GPUs not directly involved in the cross-node communication remain idle, and even active NICs are underused bacause ring attention only transfer data unidirectionally,  leaving the opposite direction unused.

To fully utilize all NICs for flexible routing, \Sys introduces a Communication Routing Layer that disaggregates logical communication paths from fixed GPU-NIC affinities. We define proxy ranks as the ranks within a node responsible for inter-node transfers. Considering an inter-node ring: let $x_1$ and $x_2$ denote the number of proxy ranks at the current node for sending and receiving, respectively. In each round of send-receives, the inter-node communication for KV activation of total size $n$ is decomposed into three-steps, as shown in Fig.~\ref{fig:routing}:
\begin{itemize}[leftmargin=*]
    \item \textbf{Workload Dispatch (Intra-node):} The source rank scatters its $n$ tokens to $x_1$ send proxy ranks using high-bandwidth intra-node links, with each proxy handling about $n/x_1$ tokens.
    \item \textbf{Inter-node Transfer (Multi-NIC):} The $x_1$ send proxy ranks deliver their data portions to the destination node, while $x_2$ receive proxy ranks simultaneously retrieve data from the source node.
    \item \textbf{Workload Combine (Intra-node):} Each of the $x_2$ receive proxy ranks forwards its $n/x_2$ tokens to the designated destination rank. 
\end{itemize}
The direct transfer cost, $b_{\text{inter}} \cdot n$, is thus optimized to:
\begin{equation}
   b_{\text{intra}}\frac{n(x1-1)}{x1} + b_{\text{inter}}\max(\frac{n}{x1}, \frac{n}{x2}) + b_{\text{intra}}\frac{n (x2-1)}{x2}
   \label{eq:disa}
\end{equation}
Given the typical 10$\times$ bandwidth gap between $b_{\text{intra}}, b_{\text{inter}}$ in modern GPU clusters, even a small number of proxy ranks can significantly reduce the inter-node communication bottleneck.

To implement this, GPUs are designated as send and receive proxy ranks in a balanced manner. GPUs already participating in ring communication act as proxy ranks for their groups, while GPUs handling local or intra-node sequences can also serve as proxies for inter-node transfer. For inter-node rings spanning multiple nodes, the number of proxy GPUs involved in communication may differ across nodes. To ensure balanced communication, the number of send proxy ranks $x_1$ at each node is set to the minimum number of GPUs assigned to the current and destination nodes; the number of receive proxy ranks $x_2$ is set analogously, based on the current and source nodes. This pairing strategy ensures one-to-one matching of senders and receivers.

The optimized inter-node communication can be scheduled to overlap with local computation and intra-node transfers. As shown in Fig.~\ref{fig:routing}.a, intra-node send-receives can proceed simultaneously with cross-node data transfer, improving hardware utilization. For ranks within a ring, their intra-node communication can also be scheduled to overlap the cross-node data transfer in Fig.~\ref{fig:routing}.b.

\subsection{Remapping Layer}
\label{remapper}
%After managing the attention imbalance, linear computation still accounts for a portion of the total overhead compared to attention computation as sequence length distribution is not large, as shown in Fig. \ref{fig:3zones}. The sequence partition patterns for attention are usually with token capacities of some ranks fully occupied, while some are deficient. However, the balanced computation and memory pattern for linear structures, such as Matmul, LayerNorm, and MoEs, is that each rank processes the same number of tokens.  As a result, the efficient partition and placement strategy for attention, which aims to balance quadratic computation, may not align with the efficient pattern for linear computation structures. 

After addressing attention imbalance, linear computation still contributes a non-negligible share of the total overhead, particularly when the sequence length distribution is highly right-skewed, with many short sequences and a few very long ones, as shown in Fig.~\ref{fig:3zones}. The partitioning strategy optimized for attention often leads to token distributions where some ranks are heavily loaded while others may be underutilized. In contrast, linear structures such as MatMul, LayerNorm, and Mixture-of-Experts (MoEs) achieve optimal efficiency when tokens are evenly distributed across ranks. Consequently, a strategy that balances quadratic attention computation may be misaligned with the optimal distribution for linear modules, leading to inefficiencies.

To bridge this gap, we introduce a Remapping Layer that dynamically adjusts the sequence distribution across ranks before and after the linear modules. Prior to linear computation, the remapping layer transforms the attention-optimized layout into a token-balanced layout. Afterwards, it performs the inverse transformation, incurring an equivalent cost, thereby aligning both attention and linear structures with their respective optimal execution patterns. %Therefore, the total remapping cost within a transformer layer is twice that of a single remapping operation. 

To minimize the communication overhead of remapping operations, we formalize the problem as an optimization task.
Let $A$ be a $d$-dimension vector representing the current token distribution across $d$ ranks in a remapping group.  
The target distribution $B$ is a $d$-dimensional vector in which each element $B_i$ equals the average number of tokens, i.e., $\sum A_i/d$. The goal is to find a transfer matrix $M \in \mathbb{R}^{d \times d}$, which minimizes the data transfer cost required to transform $A$ into $B$:
 \begin{equation}
\begin{aligned}
     \mathop{\arg \min}_{M} & \quad ||(T*M)\mathbf{1} ||_{\infty} \\
    s.t. & \sum_j(M_{ij}) = \max\{A-B, 0\}_i, \forall i;\\
         & \sum_i(M_{ij}) = \max\{B-A, 0\}_j, \forall j; \\
         & M \geq 0 ; \\
         %& T_{ij} = b_{intra} * \mathbb{I}_{intra} + b_{inter} * \mathbb{I}_{inter};
\label{eq:remapping}
\end{aligned}
 \end{equation}

\begin{figure*}[t]
  \includegraphics[width=\linewidth]{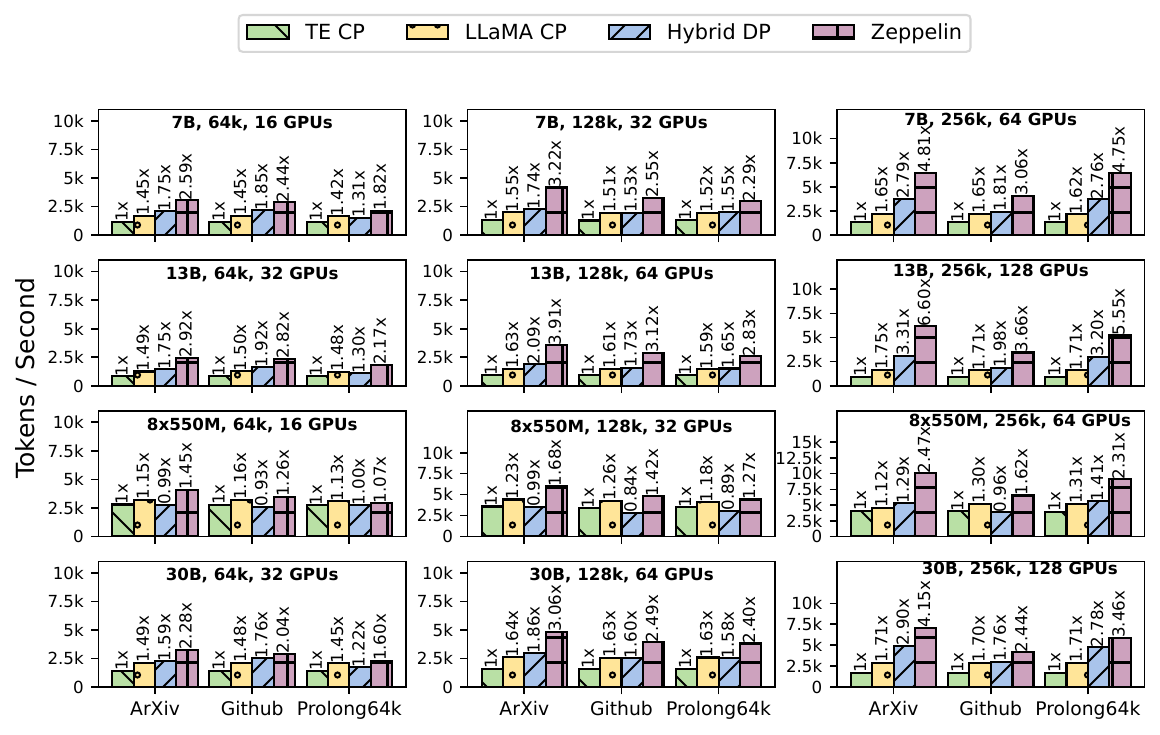}
  \caption{Throughputs for three model configurations—7B, 13B, 30B, and 8×550M—evaluated on three dataset types, across total context lengths ranging from 64k to 256k (with 4k per GPU), on Cluster A \& C.}
  \label{fig:exp1}
\end{figure*}

Here, $M_{ij}$ denotes the volume of tokens transferred from rank $i$ to rank $j$, and
$T$ is a symmetric cost matrix determined by network bandwidth: $T_{ij}=b_{\text{inter}}$ for inter-node remapping, and $T_{ij}=b_{\text{intra}}$ otherwise. 
The first constraint ensures that ranks only send surplus tokens, while the second constraint ensures that deficits are satisfied. 
The objective minimize the maximum communication cost incurred by any rank, thereby balancing the transfer overheads. This formulation corresponds to a minimum-cost flow problem and can be solved efficiently using standard solvers such as Gurobi~\cite{gurobi}. 
\iffalse
\begin{algorithm}[tb]
    \SetAlgoLined
    \SetKwInOut{Input}{Input}
    \Input{ Attention and Linear Tokens Distribution $A, B$, \\
            Devices per Node $P$, Number of Nodes $N$}
    %s = Sum($\{s_1^2, s_2^2,..., s_S^2\}$) / $N$ \; 
    Initialize $M = [0]_{PxP} $ \;
    transfer\_demand = $A-B$ \;
    %Sort $\mathcal{S}$ in descending order\;
    /* \textcolor{blue}{Within a Node} */ \\
    %Separate $\mathcal{S}$ into $\mathcal{S}_1, \mathcal{S}_2$ according to $S_2$\; 
    \For{i in range($N$)}{
        \For{j in range($P$)}{
            
        }
    }
    /* \textcolor{blue}{Cross Nodes} */ \\
    
    \SetKwInOut{Output}{Output}
    \Output{M}
    \caption{Remapping Strategy\label{alg:remapping}}
\end{algorithm}
\fi

\section{Implementation}

We implemented the core of \Sys with ~2k lines of C++/CUDA and ~6k lines of Python, extending Megatron-LM~\cite{megatron} and Transformer Engine~\cite{te}, two widely adopted open-source LLM training frameworks. The communication routing layer includes custom NCCL kernels for the three-step routing procedure, which flexibly assign inter-node send and receive ranks. These kernels are scheduled across multiple CUDA streams with a stream manager that that prioritizes kernel launches to maximize execution overlap.
The attention engine, integrated into Transformer Engine, manages the execution of three sequence queues (local, intra-node, and inter-node) on a dedicated computation stream. A global context-parallel process group is dynamically partitioned into multiple logical ring, and the engine asynchronously launches routing-layer communication kernels or intra-node send-receive operations to overlap communication with computation. The sequence partitioner and remapping layers are integrated into Megatron-LM. The sequence partitioner leverages the global batch length distribution to execute the hierarchical partitioning algorithm and generate per-rank placement planss. The remapping layer computes optimal data transfer strategies via a standard solver~\cite{gurobi} and executes them with a dynamic-shape \texttt{alltoallv} primitive that supports both forward and backward passes.

\setlength{\tabcolsep}{1.3mm}{
\begin{table}
  \centering
  \tiny
  \caption{Sequence length distribution of three datasets. Values represent the proportion of sequences within each length bin (lengths in thousands of tokens).\label{table::size}}
  \begin{tabular}{l|c|c|c|c|c|c|c|c|c}
    \hline
    Dataset & \textbf{<1} & \textbf{1-2}& \textbf{2-4} & \textbf{4-8} &\textbf{8-16} &\textbf{16-32} &\textbf{32-64} &\textbf{64-128} &\textbf{128-256}\\
    \hline
    Arxiv & 0.032 & 0.03 & 0.08 & 0.219 & 0.338 & 0.224 & 0.077 & 0 & 0\\
    \hline
    Github & 0 & 0.34 & 0.095 & 0.104 & 0.107 & 0.102 & 0.088 & 0.064 & 0.045\\
    \hline 
    Prolong64k & 0.231 & 0.042 & 0.021 & 0.012 & 0.013 & 0.008 & 0.673 & 0& 0\\
    \hline
  \end{tabular}
  
  \label{table:len_dist}
\end{table}
}

\section{Evaluation}
 \textbf{Experimental Setup} We conduct experiments on three GPU clusters: Cluster A, B, and C.
In Cluster A, each node is equipped with 8 NVIDIA A800-80G GPUs interconnected via NVSwitch, providing 400 GB/s of intra-node bandwidth. Each node also has 4 RoCE NICs, with each NIC shared by 2 GPUs, yielding an aggregate cross-node bandwidth of $4\times$200 Gb/s.
In contrast, Cluster B features nodes equipped with 8 NVIDIA H800 GPUs and 8 RoCE NICs. Cluster C nodes contain 8 H200 GPUs and 8 CX7 NICs of $8\times$400 Gb/s bandwidth, enabling significantly higher cross-node bandwidth with one-to-one GPU–NIC mapping.
All clusters run a uniform software stack: CUDA 11.8, cuDNN 8.9.6, NCCL 2.14~\cite{NCCL}, PyTorch 2.4.0~\cite{pytorch2}, FlashAttention 2.4.3~\cite{dao2022flashattentionfastmemoryefficientexact}, Megatron-LM 0.8.0rc0~\cite{megatron}, and Transformer Engine v1.8~\cite{te}.

\noindent\textbf{Model and Dataset} We use LLaMA~\cite{touvron2023llama2openfoundation} with multi-head attention as the baseline architecture, given its competitiveness and widespread adoption. Five representative configurations are evaluated: 3B, 7B, 13B, 30B dense, and 8×550M MoE.

For datasets, we evaluate on three representative long-context datasets: ArXiv~\cite{shen2024slimpajamadcunderstandingdatacombinations}, GitHub~\cite{bytescale}, and ProLong64k~\cite{64kshormixture}. Synthetic datasets are generated to match the length distributions of these benchmarks, with batch sequence lengths sampled proportionally to dataset distributions. %ArXiv exhibits a balanced distribution across a wide range of sequence lengths. GitHub shows a long-tail distribution, with a portion of sequences exceeding 64k tokens. ProLong64k is designed to evaluate long-context capabilities, consisting primarily of 64k-length sequences. 
The detailed length distributions are shown in Table~\ref{table:len_dist}.
Throughput is reported as processed tokens per second, averaged over steps 50-150.

\noindent\textbf{Baseline}
We compare \Sys against several state-of-the-art methods:
\begin{itemize}[leftmargin=*]
    \item \textbf{Transformer Engine CP~\cite{te}}: evenly splits sequences across devices and applies balanced ring attention.
    \item \textbf{LLaMA CP \cite{grattafiori2024llama3herdmodels, wlb}}: replicates the CP approach in LLaMA training, where KV activations are all-gathered across devices prior to attention computation.
    \item \textbf{Hybrid DP~\cite{bytescale}}: combines standard DP for short sequences with ring-based CP for long sequences to balance FLOPs. When short sequences exceed memory limits, they are further chunked into smaller micro-batches.
\end{itemize}

\noindent\textbf{Evaluation Scope}
We evaluate \Sys across four aspects:
$\textcircled{1}$ end-to-end performance across models, datasets, and scales;
$\textcircled{2}$ scalability and adaptability to different network architectures; $\textcircled{3}$ ablation studies of key design points; 
$\textcircled{4}$ case studies with execution timeline analysis and length distribution effects.

\subsection{End-to-End Throughput}
We evaluate end-to-end training throughput across diverse combinations of four model architectures, three datasets, and three context length scales, as shown in Fig.~\ref{fig:exp1}. \Sys consistently outperforms other SOTA baselines, achieving up to 6.60× speedup and an average speedup of 2.80× over the TE baseline.

For 7B model training, throughput improvements are consistent across different context lengths and dataset distributions.
At each context length, datasets with shorter length distributions (e.g., ArXiv) are partitioned efficiently to reduce more communication costs than skewed distributions (e.g., GitHub) with longer sequence lengths, resulting in larger speedups.
For the 13B model on Cluster A and 30B model on Cluster C, context parallel training is combined with tensor parallelism of size 2. The speedup trend mirrors the 7B case.
An interesting observation is that the speedups achieved by the 13B model are greater than those of the 7B model. This is because, in the 13B setting, two ranks within a TP group share the same NIC on Cluster A, whereas in the 7B setting (TP size 1), the workload is handled by a single rank. Consequently, bottleneck communication is eliminated with better speedups. 

For MoE models, flop cannot be accurately estimated prior to routing, which undermines Hybrid DP's flop-based token assignment and often leads to imbalanced expert computation.
At shorter context length (e.g., 64k), where expert computation dominates, the balanced LLaMA CP method achieves the highest throughput. However, as context length grows, the primary bottleneck shifts to attention computation. In this regime, \Sys's attention optimizations dominate, yielding superior performance.

\subsection{Scalability}
We evaluate the scalability of \Sys across different training scales and datasets, and also test its performance under a different GPU-NIC affinity architecture on Cluster B.

\begin{figure}[t]
  \includegraphics[width=\linewidth]{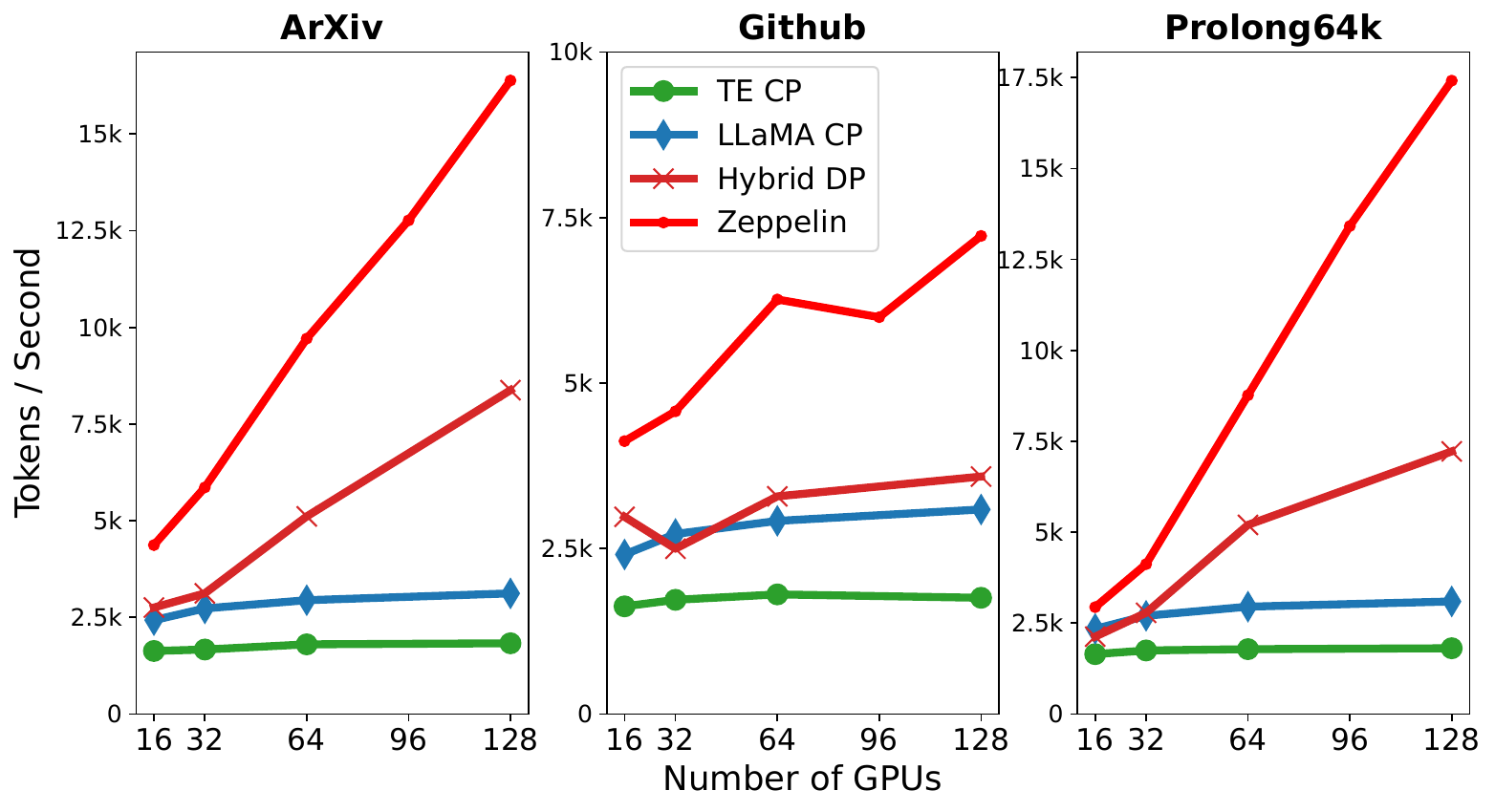}
  % \caption{Scalability of four methods with total context lengths set to \#GPUs$\times$4k for the 3B model, evaluated on Cluster A.}
  \caption{Scalability on the LLaMA 3B (Cluster A). Throughput is plotted against the number of GPUs (16-128), with context length set fixed at 4K tokens per GPU.}
  \label{fig:exp2}
\end{figure}

\begin{figure}[t]
  \includegraphics[width=\linewidth]{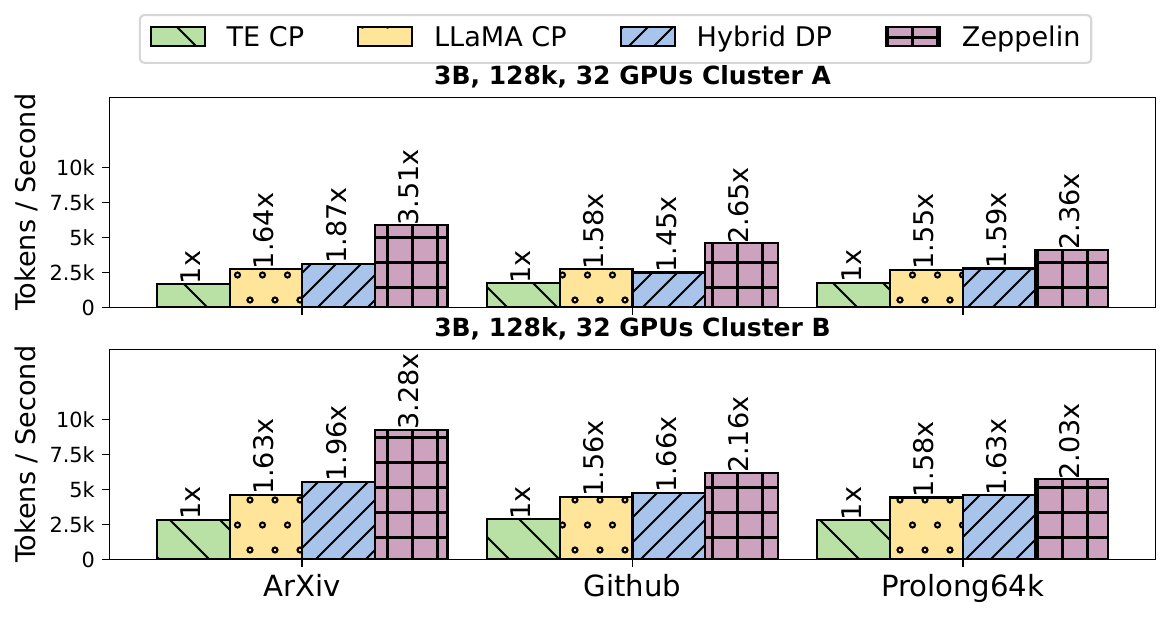}
  \caption{Speedup comparison on Cluster A \& B.}
  \label{fig:exp3}
\end{figure}

Experiments are conducted with the LLaMA 3B dense model on cluster A, using a context length of 4k tokens per GPU. 
As shown in Fig.~\ref{fig:exp2}, the TE baseline maintains nearly constant throughput across scales due to the cross-node communication bottleneck of ring attention. 
LLaMA CP achieves better performance by leveraging optimized all-gather collectives, but its communication overhead grows linearly with total sequence length.
Hybrid DP reduces communication costs using coarse-grained model-level parallelism, yet fails to capture the scaling characteristics of individual components. Consequently, it fails to outperform LLaMA CP even at small scales (16–32 GPUs), where group size remains limited.
In contrast, \Sys incorporates per-sequence parallelism and leverages the hierarchical bandwidth structure of the hardware, yielding significantly better scalability across all settings.

For the ArXiv dataset, which exhibits a balanced sequence length distribution at 64k, performance scales smoothly across all methods. In contrast, the GitHub dataset contains long sequences exceeding 64k tokens. As context length grows, such sequences become more prevalent, leading to slower throughput gains for Hybrid DP and \Sys compared to balanced datasets. For both GitHub and ProLong64k, Hybrid DP’s FLOP-balanced metric is particularly sensitive to the dominance of long sequences: they occupy most ranks and force short sequences to be split into additional micro-batches. By contrast, \Sys mitigates this bottleneck through its fine-grained, per-sequence execution and hierarchical partitioning strategy.

We further evaluate performance on Cluster B, which features a different GPU–NIC affinity architecture and stronger GPU processing capabilities. As shown in Fig.~\ref{fig:exp3}, \Sys achieves significant throughput improvements on Cluster B over all baselines. The overall throughput is higher than that of Cluster A, primarily due to the enhanced computational power of the Hopper architecture. Both clusters exhibit similar speedup trends across datasets. 
However, the relative speedup of \Sys is greater on Cluster A. This is because Cluster A has a larger computation-to-communication ratio, allowing \Sys to more effectively overlap communication costs, consistent with its design principles.

\subsection{Ablation Study}

\begin{figure}[t]
  \includegraphics[width=\linewidth]{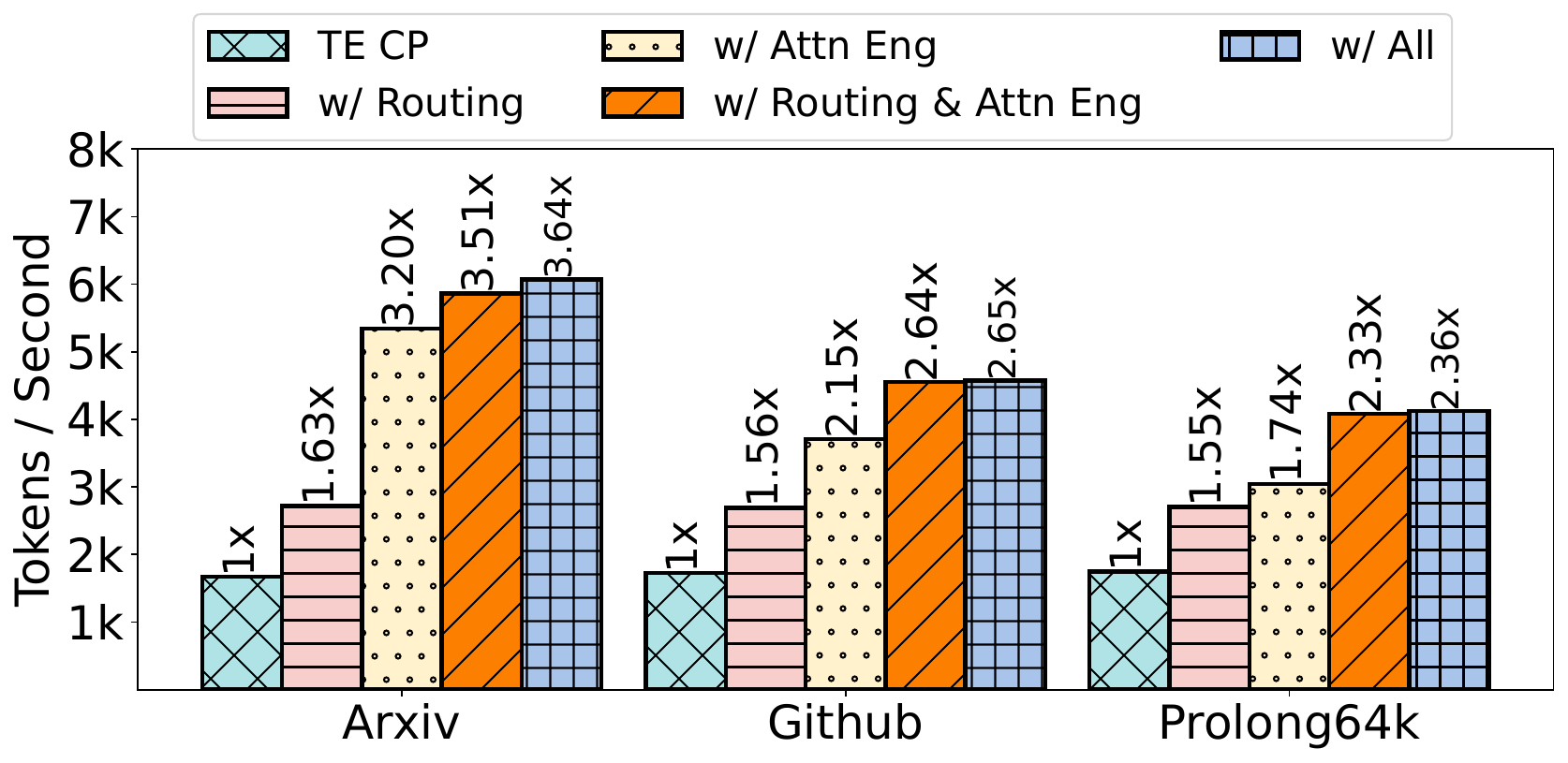}
  \caption{Ablation study on \Sys component performance. Comparison of TE CP baseline versus configurations with Routing Layer, Attention Engine, and Remapping Layer on a 3B model (32 GPUs, Cluster A) across three datasets.}
  \label{fig:exp4}
\end{figure}

\begin{figure*}
  \includegraphics[width=\linewidth]{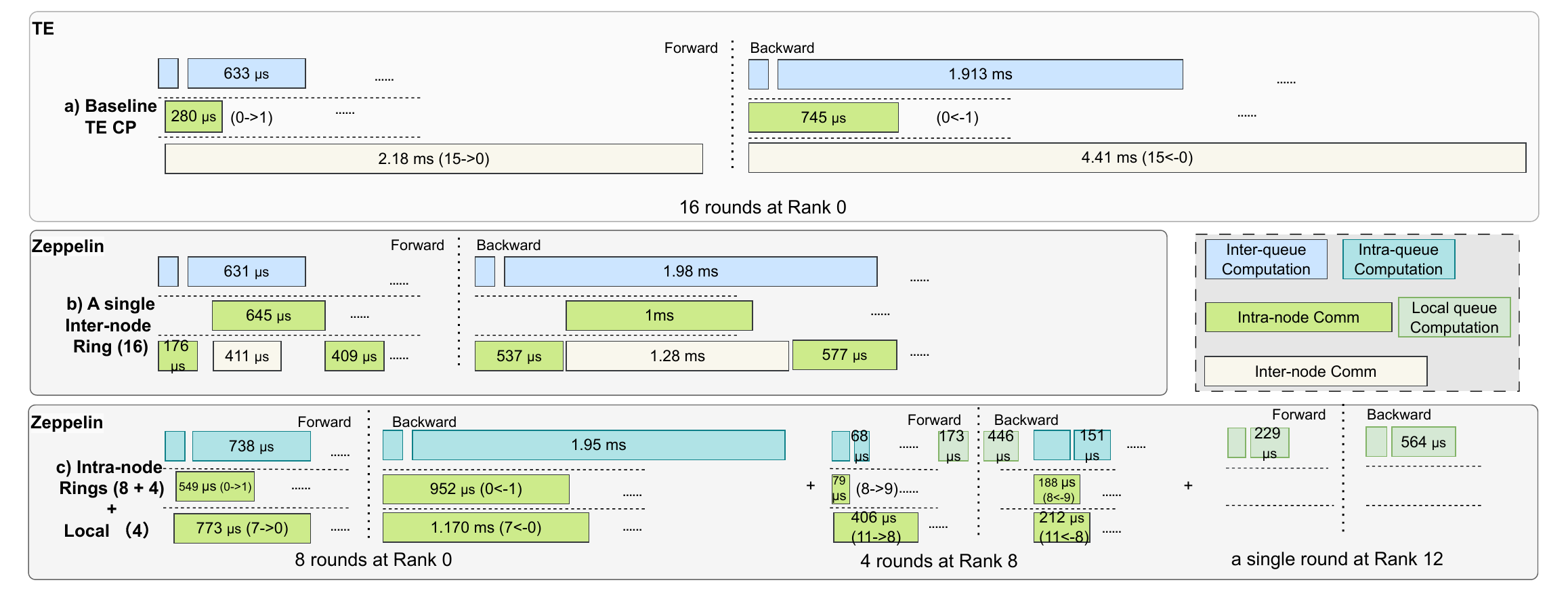}
  \caption{Forward and backward phase timelines of the attention component in a 3B model on 16 GPUs with a total context length of 64k tokens.
a) TE CP: For a single sequence of length 64k, cross-node communication between rank 0 and rank 15 becomes a bottleneck during each round of ring attention.
b) \Sys (single sequence): For a 64k-length sequence, the cross-node communication workload is split into eight chunks, enabling local dispatch, inter-node transfers, and local combination in each ring attention round.
c) \Sys (multiple sequences): For multiple sequences with a total length of 64k, sequences are distributed to separate nodes without partitioning across nodes. Within each node, sequences are further partitioned or placed with balanced computation, improving parallel efficiency.}
  \label{fig:exp5}
\end{figure*}

We evaluate the efficacy of key design components in \Sys by comparing TE CP with ablated variants of \Sys across three datasets, as shown in Fig.~\ref{fig:exp4}.
Integrating the Communication Routing Layer into the TE CP baseline enables full utilization of all NICs in a global ring. Since the total communication volume remains proportional to token count, the routing layer alone provides a consistent $\sim$1.6× speedup across datasets.
Adding efficient sequence partitioning and the attention engine, which coordinate the execution of local, intra-node, and inter-node queues, further reduces communication volume. On balanced datasets such as ArXiv, this yields up to 3.2× speedup. Combining routing with the attention engine amplifies these gains by simultaneously reducing communication volume and improving NIC utilization.
Building on attention-level partitioning, introducing the Remapping Layer balances workloads for linear modules. On right-skewed datasets such as ArXiv, remapping increases speedup from 3.51× to 3.64×. In contrast, on long-sequence–dominated datasets such as GitHub, the incremental benefit of remapping is minimal, since attention computation dominates the cost.

\subsection{Case Study}
\subsubsection{Timeline Analysis}
We profile three execution traces of a 3B model on 16 GPUs wit a total sequence length of 64k on Cluster A in Fig.~\ref{fig:exp5}. These traces illustrate how \Sys improves performance and reveal potential overheads in its implementation.
In the TE CP baseline, each round of ring attention overlaps attention computation of local Query activations and inter-node communication of KV activations for the next round. However, limited inter-node bandwidth makes inter-node communication the dominant contributor to attention overhead.

With communication routing in \Sys (Fig.~\ref{fig:exp5}.b), inter-node communication is decomposed into three steps. The local dispatch step overlaps directly with attention computation, but the inter-node transfer phase often stalls while waiting for Streaming Multiprocessor (SM) resources, since attention kernels dominate compute capacity. This creates bubbles—idle periods between communication kernels in the forward phase. Despite these bubbles, inter-node communication cost is reduced substantially from 2.18 ms to 411 $\mu$s (a reduction proportional to NIC count). The forward phase communication cost is nearly halved, decreasing from 2.18 ms to 1.3 ms. 
In the backward phase, both computation and communication roughly double in duration, but overlap improves.

When the total context consists of multiple small sequences (Fig.~\ref{fig:exp5}.c), inter-node communication is avoided by assigning sequences to separate node buckets. Intra-node communication overlaps more effectively with computation, as observed in the timeline of Rank 0. Ranks 8–11 handle both intra-node and local sequences, with backward computation executed in reverse order. For small intra-node communications, performance fluctuations appear at Rank 8, in contrast to the stable but heavier communication observed in TE CP. Nonetheless, the per-round cost is dramatically reduced from  16$\times$ (2.18 ms + 4.41 ms) = 105.44 ms in TE CP to 8$\times$ (738 $\mu$s + 1.95 ms) = 21.504 ms in \Sys.

\subsubsection{Length Distribution}

\begin{table}
  \centering
  
  \caption{ Cost distribution under two length distribution\label{table::distribution}}
  \begin{tabular}{l|c|c}
    \hline
    Components (ms) & Balanced & Skewed \\
    \hline
    Forward &  316 - 817 &   1000 - 1002\\
    \hline
    Forward Quadratic Attention & 161 - 670 & 801 - 854 \\
    \hline
    Forward Linear Modules &  93 - 97 &91 - 105 \\
    \hline
    Forward Remapping Layer& 14 - 16 & 14 - 45\\
    \hline
   Forward Sequence Partition & 3 - 10 & 3 - 12 \\
    \hline 
    Backward &  422 - 1108 & 1605 - 1608\\
    \hline
  \end{tabular}
  
\end{table}

We evaluate the cost distribution across ranks under two different input length distributions, as shown in Table~\ref{table::distribution}. Experiments are conducted on the 7B model with four nodes of Cluster C and a total context length of 128K. The Balanced distribution samples sequences from each bucket in Table~\ref{table::size}, while the Skewed distribution contains one very long sequence and several short ones.
End-to-end overhead under the Skewed distribution is higher than under the Balanced distribution, as the long sequence dominates attention computation. By contrast, remapping overheads are negligible, since our partitioning algorithm already accounts for token balance. The complexity of sequence partitioning is polynomial and incurred only once per iteration, making it negligible compared to the end-to-end training cost.

\section{Related Work}
Training LLMs efficiently at scale and enabling long-context capability have driven the development of advanced distributed learning techniques. Our work builds upon and extends prior research in parallelization strategies, long-sequence processing, and communication optimization.

\noindent\textbf{Basic Parallelisms in LLM Training} In Data Parallelism~\cite{deepspeed}, multiple workers train identical model replicas on different data subsets. ZeRO~\cite{rajbhandari2020zeromemoryoptimizationstraining} reduces memory redundancy in DP. For models too large to fit on a single device, Tensor Parallelism (TP)~\cite{megatron, megatronsp} shards weights within layers, while Pipeline Parallelism (PP)~\cite{megatron2, harlap2018pipedreamfastefficientpipeline,qi2023zerobubblepipelineparallelism, chen2025crosspipeoptimalpipelineschedules} partitions layers across devices. Hybrid approaches~\cite{kuaishouatc24, Centauri,megascale} combine these techniques to scale to extremely large models.

\noindent\textbf{Sequence Length Scaling} The growing demand for long-context processing stresses traditional parallelisms due to the quadratic cost of self-attention mechanism. FlashAttention~\cite{dao2022flashattentionfastmemoryefficientexact} reduces memory I/O overheads for single-device attention. For distributed training, multiple schemes have been proposed: DeepSpeed Ulysses~\cite{ulysses} utilizes all-to-all operations to switch between sequence- and head-wise partitions; WLB-LLM~\cite{wlb} allgathers KV cache before local attention; Ring Attention~\cite{liu2023ringattentionblockwisetransformers} adopts ring-based communication to distribute attention; Striped Attention~\cite{brandon2023stripedringattentionfasterring} and DISTFLASHATTN~\cite{li2024distflashattndistributedmemoryefficientattention} optimize load balance under causal masks; USP~\cite{usp} integrate DeepSpeed-Ulysses and Ring Attention for better scalability; and LoongTrain~\cite{gu2024loongtrainefficienttraininglongsequence} applies a double ring algorithm to reduce communication overhead.

\noindent\textbf{Variable-length Training} Fixed-length padding is inefficient as it wasted computation and memory. Packing variable-length sequences~\cite{applebucket} reduces waste but leads to complex attention masks and load imbalance in SP approaches. Transformer Engine~\cite{te} balance computation by evenly splitting sequences. FlexSP~\cite{flexsp} adapts grouping and parallelism dynamically. HotSPa~\cite{hotswitch} mitigates skewness via hot switching among parallel strategies. ByteScale~\cite{bytescale} reduces redundant communication for short sequences using hybrid data parallel approaches.

\section{Conclusion}
In this paper, we proposed \Sys, a novel system design engineered to address the severe imbalance issues encountered during data-parallel training of LLMs with long and variable-length data. \Sys enhances training performance through three key components: an efficient attention engine, a flexible communication routing layer, and a remapping layer. Comprehensive evaluations demonstrate that \Sys achieves an impressive average speedup of 2.80$\times$ over state-of-the-art methods.

% In this paper, we propose a novel system design \Sys to alleviate the severe imbalance in long and variable-length data parallel large model training. \Sys improves the training performance with three key design points: an efficient attention engine, a flexible communication routing layer, and a remapping layer. Comprehensive evaluations demonstrate that \Sys achieves 2.80$\times$ average speedup over state-of-the-art methods.

\normalem
\bibliographystyle{ACM-Reference-Format}
\bibliography{reference}


\begin{thebibliography}{58}

%%% the Web address.

\ifx \showCODEN    \undefined \def \showCODEN     #1{\unskip}     \fi
\ifx \showISBNx    \undefined \def \showISBNx     #1{\unskip}     \fi
\ifx \showISBNxiii \undefined \def \showISBNxiii  #1{\unskip}     \fi
\ifx \showISSN     \undefined \def \showISSN      #1{\unskip}     \fi
\ifx \showLCCN     \undefined \def \showLCCN      #1{\unskip}     \fi
\ifx \shownote     \undefined \def \shownote      #1{#1}          \fi
\ifx \showarticletitle \undefined \def \showarticletitle #1{#1}   \fi
\ifx \showURL      \undefined \def \showURL       {\relax}        \fi
% invisible to TeX
\providecommand\bibfield[2]{#2}
\providecommand\bibinfo[2]{#2}
\providecommand\natexlab[1]{#1}
\providecommand\showeprint[2][]{arXiv:#2}

\bibitem[gur(2025)]%
        {gurobi}
 \bibinfo{year}{2025}\natexlab{}.
\newblock \bibinfo{title}{GUROBI}.
\newblock
\urldef\tempurl%
\url{https://docs.gurobi.com/projects/optimizer/en/current/index.html}
\showURL{%
\tempurl}


\bibitem[Ansel et~al\mbox{.}(2024)]%
        {pytorch2}
\bibfield{author}{\bibinfo{person}{Jason Ansel}, \bibinfo{person}{Edward Yang}, \bibinfo{person}{Horace He}, \bibinfo{person}{Natalia Gimelshein}, \bibinfo{person}{Animesh Jain}, \bibinfo{person}{Michael Voznesensky}, \bibinfo{person}{Bin Bao}, \bibinfo{person}{Peter Bell}, \bibinfo{person}{David Berard}, \bibinfo{person}{Evgeni Burovski}, \bibinfo{person}{Geeta Chauhan}, \bibinfo{person}{Anjali Chourdia}, \bibinfo{person}{Will Constable}, \bibinfo{person}{Alban Desmaison}, \bibinfo{person}{Zachary DeVito}, \bibinfo{person}{Elias Ellison}, \bibinfo{person}{Will Feng}, \bibinfo{person}{Jiong Gong}, \bibinfo{person}{Michael Gschwind}, \bibinfo{person}{Brian Hirsh}, \bibinfo{person}{Sherlock Huang}, \bibinfo{person}{Kshiteej Kalambarkar}, \bibinfo{person}{Laurent Kirsch}, \bibinfo{person}{Michael Lazos}, \bibinfo{person}{Mario Lezcano}, \bibinfo{person}{Yanbo Liang}, \bibinfo{person}{Jason Liang}, \bibinfo{person}{Yinghai Lu}, \bibinfo{person}{C.~K. Luk}, \bibinfo{person}{Bert Maher}, \bibinfo{person}{Yunjie
  Pan}, \bibinfo{person}{Christian Puhrsch}, \bibinfo{person}{Matthias Reso}, \bibinfo{person}{Mark Saroufim}, \bibinfo{person}{Marcos~Yukio Siraichi}, \bibinfo{person}{Helen Suk}, \bibinfo{person}{Shunting Zhang}, \bibinfo{person}{Michael Suo}, \bibinfo{person}{Phil Tillet}, \bibinfo{person}{Xu Zhao}, \bibinfo{person}{Eikan Wang}, \bibinfo{person}{Keren Zhou}, \bibinfo{person}{Richard Zou}, \bibinfo{person}{Xiaodong Wang}, \bibinfo{person}{Ajit Mathews}, \bibinfo{person}{William Wen}, \bibinfo{person}{Gregory Chanan}, \bibinfo{person}{Peng Wu}, {and} \bibinfo{person}{Soumith Chintala}.} \bibinfo{year}{2024}\natexlab{}.
\newblock \showarticletitle{PyTorch 2: Faster Machine Learning Through Dynamic Python Bytecode Transformation and Graph Compilation}. In \bibinfo{booktitle}{\emph{Proceedings of the 29th ACM International Conference on Architectural Support for Programming Languages and Operating Systems, Volume 2}} (La Jolla, CA, USA) \emph{(\bibinfo{series}{ASPLOS '24})}. \bibinfo{publisher}{Association for Computing Machinery}, \bibinfo{address}{New York, NY, USA}, \bibinfo{pages}{929–947}.
\newblock
\showISBNx{9798400703850}
\href{https://doi.org/10.1145/3620665.3640366}{doi:\nolinkurl{10.1145/3620665.3640366}}


\bibitem[Bertsch et~al\mbox{.}(2023)]%
        {bertsch2023unlimiformerlong}
\bibfield{author}{\bibinfo{person}{Amanda Bertsch}, \bibinfo{person}{Uri Alon}, \bibinfo{person}{Graham Neubig}, {and} \bibinfo{person}{Matthew~R. Gormley}.} \bibinfo{year}{2023}\natexlab{}.
\newblock \showarticletitle{Unlimiformer: Long-Range Transformers with Unlimited Length Input}. In \bibinfo{booktitle}{\emph{Thirty-seventh Conference on Neural Information Processing Systems}}.
\newblock
\urldef\tempurl%
\url{https://openreview.net/forum?id=lJWUJWLCJo}
\showURL{%
\tempurl}


\bibitem[Brandon et~al\mbox{.}(2023)]%
        {brandon2023stripedringattentionfasterring}
\bibfield{author}{\bibinfo{person}{William Brandon}, \bibinfo{person}{Aniruddha Nrusimha}, \bibinfo{person}{Kevin Qian}, \bibinfo{person}{Zachary Ankner}, \bibinfo{person}{Tian Jin}, \bibinfo{person}{Zhiye Song}, {and} \bibinfo{person}{Jonathan Ragan-Kelley}.} \bibinfo{year}{2023}\natexlab{}.
\newblock \bibinfo{title}{Striped Attention: Faster Ring Attention for Causal Transformers}.
\newblock
\showeprint[arxiv]{2311.09431}~[cs.LG]
\urldef\tempurl%
\url{https://arxiv.org/abs/2311.09431}
\showURL{%
\tempurl}


\bibitem[Brown et~al\mbox{.}(2020)]%
        {gpt}
\bibfield{author}{\bibinfo{person}{Tom~B. Brown}, \bibinfo{person}{Benjamin Mann}, \bibinfo{person}{Nick Ryder}, \bibinfo{person}{Melanie Subbiah}, \bibinfo{person}{Jared Kaplan}, \bibinfo{person}{Prafulla Dhariwal}, \bibinfo{person}{Arvind Neelakantan}, \bibinfo{person}{Pranav Shyam}, \bibinfo{person}{Girish Sastry}, \bibinfo{person}{Amanda Askell}, \bibinfo{person}{Sandhini Agarwal}, \bibinfo{person}{Ariel Herbert{-}Voss}, \bibinfo{person}{Gretchen Krueger}, \bibinfo{person}{Tom Henighan}, \bibinfo{person}{Rewon Child}, \bibinfo{person}{Aditya Ramesh}, \bibinfo{person}{Daniel~M. Ziegler}, \bibinfo{person}{Jeffrey Wu}, \bibinfo{person}{Clemens Winter}, \bibinfo{person}{Christopher Hesse}, \bibinfo{person}{Mark Chen}, \bibinfo{person}{Eric Sigler}, \bibinfo{person}{Mateusz Litwin}, \bibinfo{person}{Scott Gray}, \bibinfo{person}{Benjamin Chess}, \bibinfo{person}{Jack Clark}, \bibinfo{person}{Christopher Berner}, \bibinfo{person}{Sam McCandlish}, \bibinfo{person}{Alec Radford}, \bibinfo{person}{Ilya Sutskever},
  {and} \bibinfo{person}{Dario Amodei}.} \bibinfo{year}{2020}\natexlab{}.
\newblock \showarticletitle{Language Models are Few-Shot Learners}.
\newblock \bibinfo{journal}{\emph{CoRR}}  \bibinfo{volume}{abs/2005.14165} (\bibinfo{year}{2020}).
\newblock
\showeprint[arXiv]{2005.14165}
\urldef\tempurl%
\url{https://arxiv.org/abs/2005.14165}
\showURL{%
\tempurl}


\bibitem[Chen et~al\mbox{.}(2024)]%
        {Centauri}
\bibfield{author}{\bibinfo{person}{Chang Chen}, \bibinfo{person}{Xiuhong Li}, \bibinfo{person}{Qianchao Zhu}, \bibinfo{person}{Jiangfei Duan}, \bibinfo{person}{Peng Sun}, \bibinfo{person}{Xingcheng Zhang}, {and} \bibinfo{person}{Chao Yang}.} \bibinfo{year}{2024}\natexlab{}.
\newblock \showarticletitle{Centauri: Enabling Efficient Scheduling for Communication-Computation Overlap in Large Model Training via Communication Partitioning}. In \bibinfo{booktitle}{\emph{Proceedings of the 29th ACM International Conference on Architectural Support for Programming Languages and Operating Systems, Volume 3}} (La Jolla, CA, USA) \emph{(\bibinfo{series}{ASPLOS '24})}. \bibinfo{publisher}{Association for Computing Machinery}, \bibinfo{address}{New York, NY, USA}, \bibinfo{pages}{178–191}.
\newblock
\showISBNx{9798400703867}
\href{https://doi.org/10.1145/3620666.3651379}{doi:\nolinkurl{10.1145/3620666.3651379}}


\bibitem[Chen et~al\mbox{.}(2025a)]%
        {mixture2}
\bibfield{author}{\bibinfo{person}{Mayee~F. Chen}, \bibinfo{person}{Michael~Y. Hu}, \bibinfo{person}{Nicholas Lourie}, \bibinfo{person}{Kyunghyun Cho}, {and} \bibinfo{person}{Christopher Ré}.} \bibinfo{year}{2025}\natexlab{a}.
\newblock \bibinfo{title}{Aioli: A Unified Optimization Framework for Language Model Data Mixing}.
\newblock
\showeprint[arxiv]{2411.05735}~[cs.LG]
\urldef\tempurl%
\url{https://arxiv.org/abs/2411.05735}
\showURL{%
\tempurl}


\bibitem[Chen et~al\mbox{.}(2025b)]%
        {chen2025crosspipeoptimalpipelineschedules}
\bibfield{author}{\bibinfo{person}{Tiancheng Chen}, \bibinfo{person}{Ales Kubicek}, \bibinfo{person}{Langwen Huang}, {and} \bibinfo{person}{Torsten Hoefler}.} \bibinfo{year}{2025}\natexlab{b}.
\newblock \bibinfo{title}{CrossPipe: Towards Optimal Pipeline Schedules for Cross-Datacenter Training}.
\newblock
\showeprint[arxiv]{2507.00217}~[cs.DC]
\urldef\tempurl%
\url{https://arxiv.org/abs/2507.00217}
\showURL{%
\tempurl}


\bibitem[Chen et~al\mbox{.}(2025c)]%
        {chen2025longvila}
\bibfield{author}{\bibinfo{person}{Yukang Chen}, \bibinfo{person}{Fuzhao Xue}, \bibinfo{person}{Dacheng Li}, \bibinfo{person}{Qinghao Hu}, \bibinfo{person}{Ligeng Zhu}, \bibinfo{person}{Xiuyu Li}, \bibinfo{person}{Yunhao Fang}, \bibinfo{person}{Haotian Tang}, \bibinfo{person}{Shang Yang}, \bibinfo{person}{Zhijian Liu}, \bibinfo{person}{Yihui He}, \bibinfo{person}{Hongxu Yin}, \bibinfo{person}{Pavlo Molchanov}, \bibinfo{person}{Jan Kautz}, \bibinfo{person}{Linxi Fan}, \bibinfo{person}{Yuke Zhu}, \bibinfo{person}{Yao Lu}, {and} \bibinfo{person}{Song Han}.} \bibinfo{year}{2025}\natexlab{c}.
\newblock \showarticletitle{Long{VILA}: Scaling Long-Context Visual Language Models for Long Videos}. In \bibinfo{booktitle}{\emph{The Thirteenth International Conference on Learning Representations}}.
\newblock
\urldef\tempurl%
\url{https://openreview.net/forum?id=wCXAlfvCy6}
\showURL{%
\tempurl}


\bibitem[Dao et~al\mbox{.}(2022)]%
        {dao2022flashattentionfastmemoryefficientexact}
\bibfield{author}{\bibinfo{person}{Tri Dao}, \bibinfo{person}{Daniel~Y. Fu}, \bibinfo{person}{Stefano Ermon}, \bibinfo{person}{Atri Rudra}, {and} \bibinfo{person}{Christopher Ré}.} \bibinfo{year}{2022}\natexlab{}.
\newblock \bibinfo{title}{FlashAttention: Fast and Memory-Efficient Exact Attention with IO-Awareness}.
\newblock
\showeprint[arxiv]{2205.14135}~[cs.LG]
\urldef\tempurl%
\url{https://arxiv.org/abs/2205.14135}
\showURL{%
\tempurl}


\bibitem[DeepSeek-AI(2024)]%
        {deepseekai2024deepseekv2strongeconomicalefficient}
\bibfield{author}{\bibinfo{person}{DeepSeek-AI}.} \bibinfo{year}{2024}\natexlab{}.
\newblock \bibinfo{title}{DeepSeek-V2: A Strong, Economical, and Efficient Mixture-of-Experts Language Model}.
\newblock
\showeprint[arxiv]{2405.04434}~[cs.CL]
\urldef\tempurl%
\url{https://arxiv.org/abs/2405.04434}
\showURL{%
\tempurl}


\bibitem[DeepSeek-AI(2025)]%
        {deepseekai2025deepseekv3technicalreport}
\bibfield{author}{\bibinfo{person}{DeepSeek-AI}.} \bibinfo{year}{2025}\natexlab{}.
\newblock \bibinfo{title}{DeepSeek-V3 Technical Report}.
\newblock
\showeprint[arxiv]{2412.19437}~[cs.CL]
\urldef\tempurl%
\url{https://arxiv.org/abs/2412.19437}
\showURL{%
\tempurl}


\bibitem[Dehghani et~al\mbox{.}(2023)]%
        {deepmindpack}
\bibfield{author}{\bibinfo{person}{Mostafa Dehghani}, \bibinfo{person}{Basil Mustafa}, \bibinfo{person}{Josip Djolonga}, \bibinfo{person}{Jonathan Heek}, \bibinfo{person}{Matthias Minderer}, \bibinfo{person}{Mathilde Caron}, \bibinfo{person}{Andreas Steiner}, \bibinfo{person}{Joan Puigcerver}, \bibinfo{person}{Robert Geirhos}, \bibinfo{person}{Ibrahim~M Alabdulmohsin}, \bibinfo{person}{Avital Oliver}, \bibinfo{person}{Piotr Padlewski}, \bibinfo{person}{Alexey Gritsenko}, \bibinfo{person}{Mario Lucic}, {and} \bibinfo{person}{Neil Houlsby}.} \bibinfo{year}{2023}\natexlab{}.
\newblock \showarticletitle{Patch n’ Pack: NaViT, a Vision Transformer for any Aspect Ratio and Resolution}. In \bibinfo{booktitle}{\emph{Advances in Neural Information Processing Systems}}, \bibfield{editor}{\bibinfo{person}{A.~Oh}, \bibinfo{person}{T.~Naumann}, \bibinfo{person}{A.~Globerson}, \bibinfo{person}{K.~Saenko}, \bibinfo{person}{M.~Hardt}, {and} \bibinfo{person}{S.~Levine}} (Eds.), Vol.~\bibinfo{volume}{36}. \bibinfo{publisher}{Curran Associates, Inc.}, \bibinfo{pages}{2252--2274}.
\newblock
\urldef\tempurl%
\url{https://proceedings.neurips.cc/paper_files/paper/2023/file/06ea400b9b7cfce6428ec27a371632eb-Paper-Conference.pdf}
\showURL{%
\tempurl}


\bibitem[Ding et~al\mbox{.}(2024)]%
        {ding2024fewertruncationspackingimprovelanguage}
\bibfield{author}{\bibinfo{person}{Hantian Ding}, \bibinfo{person}{Zijian Wang}, \bibinfo{person}{Giovanni Paolini}, \bibinfo{person}{Varun Kumar}, \bibinfo{person}{Anoop Deoras}, \bibinfo{person}{Dan Roth}, {and} \bibinfo{person}{Stefano Soatto}.} \bibinfo{year}{2024}\natexlab{}.
\newblock \bibinfo{title}{Fewer Truncations Improve Language Modeling}.
\newblock
\showeprint[arxiv]{2404.10830}~[cs.CL]
\urldef\tempurl%
\url{https://arxiv.org/abs/2404.10830}
\showURL{%
\tempurl}


\bibitem[Dwarampudi and Reddy(2019)]%
        {padding}
\bibfield{author}{\bibinfo{person}{Mahidhar Dwarampudi} {and} \bibinfo{person}{N~V~Subba Reddy}.} \bibinfo{year}{2019}\natexlab{}.
\newblock \bibinfo{title}{Effects of padding on LSTMs and CNNs}.
\newblock
\showeprint[arxiv]{1903.07288}~[cs.LG]
\urldef\tempurl%
\url{https://arxiv.org/abs/1903.07288}
\showURL{%
\tempurl}


\bibitem[Fang and Zhao(2024)]%
        {usp}
\bibfield{author}{\bibinfo{person}{Jiarui Fang} {and} \bibinfo{person}{Shangchun Zhao}.} \bibinfo{year}{2024}\natexlab{}.
\newblock \bibinfo{title}{USP: A Unified Sequence Parallelism Approach for Long Context Generative AI}.
\newblock
\showeprint[arxiv]{2405.07719}~[cs.LG]
\urldef\tempurl%
\url{https://arxiv.org/abs/2405.07719}
\showURL{%
\tempurl}


\bibitem[Fu et~al\mbox{.}(2024)]%
        {mixture1fuyao}
\bibfield{author}{\bibinfo{person}{Yao Fu}, \bibinfo{person}{Rameswar Panda}, \bibinfo{person}{Xinyao Niu}, \bibinfo{person}{Xiang Yue}, \bibinfo{person}{Hannaneh Hajishirzi}, \bibinfo{person}{Yoon Kim}, {and} \bibinfo{person}{Hao Peng}.} \bibinfo{year}{2024}\natexlab{}.
\newblock \showarticletitle{Data engineering for scaling language models to 128K context}. In \bibinfo{booktitle}{\emph{Proceedings of the 41st International Conference on Machine Learning}} (Vienna, Austria) \emph{(\bibinfo{series}{ICML'24})}. \bibinfo{publisher}{JMLR.org}, Article \bibinfo{articleno}{564}, \bibinfo{numpages}{10}~pages.
\newblock


\bibitem[Gao et~al\mbox{.}(2025)]%
        {64kshormixture}
\bibfield{author}{\bibinfo{person}{Tianyu Gao}, \bibinfo{person}{Alexander Wettig}, \bibinfo{person}{Howard Yen}, {and} \bibinfo{person}{Danqi Chen}.} \bibinfo{year}{2025}\natexlab{}.
\newblock \bibinfo{title}{How to Train Long-Context Language Models (Effectively)}.
\newblock
\showeprint[arxiv]{2410.02660}~[cs.CL]
\urldef\tempurl%
\url{https://arxiv.org/abs/2410.02660}
\showURL{%
\tempurl}


\bibitem[Ge et~al\mbox{.}(2025)]%
        {bytescale}
\bibfield{author}{\bibinfo{person}{Hao Ge}, \bibinfo{person}{Junda Feng}, \bibinfo{person}{Qi Huang}, \bibinfo{person}{Fangcheng Fu}, \bibinfo{person}{Xiaonan Nie}, \bibinfo{person}{Lei Zuo}, \bibinfo{person}{Haibin Lin}, \bibinfo{person}{Bin Cui}, {and} \bibinfo{person}{Xin Liu}.} \bibinfo{year}{2025}\natexlab{}.
\newblock \bibinfo{title}{ByteScale: Efficient Scaling of LLM Training with a 2048K Context Length on More Than 12,000 GPUs}.
\newblock
\showeprint[arxiv]{2502.21231}~[cs.DC]
\urldef\tempurl%
\url{https://arxiv.org/abs/2502.21231}
\showURL{%
\tempurl}


\bibitem[Ge et~al\mbox{.}(2024)]%
        {hotswitch}
\bibfield{author}{\bibinfo{person}{Hao Ge}, \bibinfo{person}{Fangcheng Fu}, \bibinfo{person}{Haoyang Li}, \bibinfo{person}{Xuanyu Wang}, \bibinfo{person}{Sheng Lin}, \bibinfo{person}{Yujie Wang}, \bibinfo{person}{Xiaonan Nie}, \bibinfo{person}{Hailin Zhang}, \bibinfo{person}{Xupeng Miao}, {and} \bibinfo{person}{Bin Cui}.} \bibinfo{year}{2024}\natexlab{}.
\newblock \showarticletitle{Enabling Parallelism Hot Switching for Efficient Training of Large Language Models}. In \bibinfo{booktitle}{\emph{Proceedings of the ACM SIGOPS 30th Symposium on Operating Systems Principles}} (Austin, TX, USA) \emph{(\bibinfo{series}{SOSP '24})}. \bibinfo{publisher}{Association for Computing Machinery}, \bibinfo{address}{New York, NY, USA}, \bibinfo{pages}{178–194}.
\newblock
\showISBNx{9798400712517}
\href{https://doi.org/10.1145/3694715.3695969}{doi:\nolinkurl{10.1145/3694715.3695969}}


\bibitem[Google(2025)]%
        {gemini}
\bibfield{author}{\bibinfo{person}{Google}.} \bibinfo{year}{2025}\natexlab{}.
\newblock \bibinfo{title}{Gemini 2.5: Our most intelligent AI model}.
\newblock
\urldef\tempurl%
\url{https://blog.google/technology/google-deepmind/gemini-model-thinking-updates-march-2025/#gemini-2-5-thinking}
\showURL{%
\tempurl}


\bibitem[Goyal et~al\mbox{.}(2018)]%
        {goyal2018accuratelargeminibatchsgd}
\bibfield{author}{\bibinfo{person}{Priya Goyal}, \bibinfo{person}{Piotr Dollár}, \bibinfo{person}{Ross Girshick}, \bibinfo{person}{Pieter Noordhuis}, \bibinfo{person}{Lukasz Wesolowski}, \bibinfo{person}{Aapo Kyrola}, \bibinfo{person}{Andrew Tulloch}, \bibinfo{person}{Yangqing Jia}, {and} \bibinfo{person}{Kaiming He}.} \bibinfo{year}{2018}\natexlab{}.
\newblock \bibinfo{title}{Accurate, Large Minibatch SGD: Training ImageNet in 1 Hour}.
\newblock
\showeprint[arxiv]{1706.02677}~[cs.CV]
\urldef\tempurl%
\url{https://arxiv.org/abs/1706.02677}
\showURL{%
\tempurl}


\bibitem[Gu et~al\mbox{.}(2024)]%
        {gu2024loongtrainefficienttraininglongsequence}
\bibfield{author}{\bibinfo{person}{Diandian Gu}, \bibinfo{person}{Peng Sun}, \bibinfo{person}{Qinghao Hu}, \bibinfo{person}{Ting Huang}, \bibinfo{person}{Xun Chen}, \bibinfo{person}{Yingtong Xiong}, \bibinfo{person}{Guoteng Wang}, \bibinfo{person}{Qiaoling Chen}, \bibinfo{person}{Shangchun Zhao}, \bibinfo{person}{Jiarui Fang}, \bibinfo{person}{Yonggang Wen}, \bibinfo{person}{Tianwei Zhang}, \bibinfo{person}{Xin Jin}, {and} \bibinfo{person}{Xuanzhe Liu}.} \bibinfo{year}{2024}\natexlab{}.
\newblock \bibinfo{title}{LoongTrain: Efficient Training of Long-Sequence LLMs with Head-Context Parallelism}.
\newblock
\showeprint[arxiv]{2406.18485}~[cs.DC]
\urldef\tempurl%
\url{https://arxiv.org/abs/2406.18485}
\showURL{%
\tempurl}


\bibitem[Harlap et~al\mbox{.}(2018)]%
        {harlap2018pipedreamfastefficientpipeline}
\bibfield{author}{\bibinfo{person}{Aaron Harlap}, \bibinfo{person}{Deepak Narayanan}, \bibinfo{person}{Amar Phanishayee}, \bibinfo{person}{Vivek Seshadri}, \bibinfo{person}{Nikhil Devanur}, \bibinfo{person}{Greg Ganger}, {and} \bibinfo{person}{Phil Gibbons}.} \bibinfo{year}{2018}\natexlab{}.
\newblock \bibinfo{title}{PipeDream: Fast and Efficient Pipeline Parallel DNN Training}.
\newblock
\showeprint[arxiv]{1806.03377}~[cs.DC]
\urldef\tempurl%
\url{https://arxiv.org/abs/1806.03377}
\showURL{%
\tempurl}


\bibitem[huggingface(2025)]%
        {huggingface}
\bibfield{author}{\bibinfo{person}{huggingface}.} \bibinfo{year}{2025}\natexlab{}.
\newblock \bibinfo{title}{Padding and truncation}.
\newblock
\urldef\tempurl%
\url{https://huggingface.co/docs/transformers/pad_truncation}
\showURL{%
\tempurl}


\bibitem[Jacobs et~al\mbox{.}(2024)]%
        {ulysses}
\bibfield{author}{\bibinfo{person}{Sam~Ade Jacobs}, \bibinfo{person}{Masahiro Tanaka}, \bibinfo{person}{Chengming Zhang}, \bibinfo{person}{Minjia Zhang}, \bibinfo{person}{Reza~Yazdani Aminadabi}, \bibinfo{person}{Shuaiwen~Leon Song}, \bibinfo{person}{Samyam Rajbhandari}, {and} \bibinfo{person}{Yuxiong He}.} \bibinfo{year}{2024}\natexlab{}.
\newblock \showarticletitle{System Optimizations for Enabling Training of Extreme Long Sequence Transformer Models}. In \bibinfo{booktitle}{\emph{Proceedings of the 43rd ACM Symposium on Principles of Distributed Computing}} (Nantes, France) \emph{(\bibinfo{series}{PODC '24})}. \bibinfo{publisher}{Association for Computing Machinery}, \bibinfo{address}{New York, NY, USA}, \bibinfo{pages}{121–130}.
\newblock
\showISBNx{9798400706684}
\href{https://doi.org/10.1145/3662158.3662806}{doi:\nolinkurl{10.1145/3662158.3662806}}


\bibitem[Jiang et~al\mbox{.}(2024)]%
        {megascale}
\bibfield{author}{\bibinfo{person}{Ziheng Jiang}, \bibinfo{person}{Haibin Lin}, \bibinfo{person}{Yinmin Zhong}, \bibinfo{person}{Qi Huang}, \bibinfo{person}{Yangrui Chen}, \bibinfo{person}{Zhi Zhang}, \bibinfo{person}{Yanghua Peng}, \bibinfo{person}{Xiang Li}, \bibinfo{person}{Cong Xie}, \bibinfo{person}{Shibiao Nong}, \bibinfo{person}{Yulu Jia}, \bibinfo{person}{Sun He}, \bibinfo{person}{Hongmin Chen}, \bibinfo{person}{Zhihao Bai}, \bibinfo{person}{Qi Hou}, \bibinfo{person}{Shipeng Yan}, \bibinfo{person}{Ding Zhou}, \bibinfo{person}{Yiyao Sheng}, \bibinfo{person}{Zhuo Jiang}, \bibinfo{person}{Haohan Xu}, \bibinfo{person}{Haoran Wei}, \bibinfo{person}{Zhang Zhang}, \bibinfo{person}{Pengfei Nie}, \bibinfo{person}{Leqi Zou}, \bibinfo{person}{Sida Zhao}, \bibinfo{person}{Liang Xiang}, \bibinfo{person}{Zherui Liu}, \bibinfo{person}{Zhe Li}, \bibinfo{person}{Xiaoying Jia}, \bibinfo{person}{Jianxi Ye}, \bibinfo{person}{Xin Jin}, {and} \bibinfo{person}{Xin Liu}.} \bibinfo{year}{2024}\natexlab{}.
\newblock \bibinfo{title}{MegaScale: Scaling Large Language Model Training to More Than 10,000 GPUs}.
\newblock
\showeprint[arxiv]{2402.15627}~[cs.LG]
\urldef\tempurl%
\url{https://arxiv.org/abs/2402.15627}
\showURL{%
\tempurl}


\bibitem[Kocetkov et~al\mbox{.}(2022)]%
        {kocetkov2022stack3tbpermissively}
\bibfield{author}{\bibinfo{person}{Denis Kocetkov}, \bibinfo{person}{Raymond Li}, \bibinfo{person}{Loubna~Ben Allal}, \bibinfo{person}{Jia Li}, \bibinfo{person}{Chenghao Mou}, \bibinfo{person}{Carlos~Muñoz Ferrandis}, \bibinfo{person}{Yacine Jernite}, \bibinfo{person}{Margaret Mitchell}, \bibinfo{person}{Sean Hughes}, \bibinfo{person}{Thomas Wolf}, \bibinfo{person}{Dzmitry Bahdanau}, \bibinfo{person}{Leandro von Werra}, {and} \bibinfo{person}{Harm de Vries}.} \bibinfo{year}{2022}\natexlab{}.
\newblock \bibinfo{title}{The Stack: 3 TB of permissively licensed source code}.
\newblock
\showeprint[arxiv]{2211.15533}~[cs.CL]
\urldef\tempurl%
\url{https://arxiv.org/abs/2211.15533}
\showURL{%
\tempurl}


\bibitem[Korthikanti et~al\mbox{.}(2022)]%
        {megatronsp}
\bibfield{author}{\bibinfo{person}{Vijay Korthikanti}, \bibinfo{person}{Jared Casper}, \bibinfo{person}{Sangkug Lym}, \bibinfo{person}{Lawrence McAfee}, \bibinfo{person}{Michael Andersch}, \bibinfo{person}{Mohammad Shoeybi}, {and} \bibinfo{person}{Bryan Catanzaro}.} \bibinfo{year}{2022}\natexlab{}.
\newblock \bibinfo{title}{Reducing Activation Recomputation in Large Transformer Models}.
\newblock
\showeprint[arxiv]{2205.05198}~[cs.LG]
\urldef\tempurl%
\url{https://arxiv.org/abs/2205.05198}
\showURL{%
\tempurl}


\bibitem[Levy et~al\mbox{.}(2024)]%
        {mixture3}
\bibfield{author}{\bibinfo{person}{Mosh Levy}, \bibinfo{person}{Alon Jacoby}, {and} \bibinfo{person}{Yoav Goldberg}.} \bibinfo{year}{2024}\natexlab{}.
\newblock \showarticletitle{Same Task, More Tokens: the Impact of Input Length on the Reasoning Performance of Large Language Models}. In \bibinfo{booktitle}{\emph{Proceedings of the 62nd Annual Meeting of the Association for Computational Linguistics (Volume 1: Long Papers)}}, \bibfield{editor}{\bibinfo{person}{Lun-Wei Ku}, \bibinfo{person}{Andre Martins}, {and} \bibinfo{person}{Vivek Srikumar}} (Eds.). \bibinfo{publisher}{Association for Computational Linguistics}, \bibinfo{address}{Bangkok, Thailand}, \bibinfo{pages}{15339--15353}.
\newblock
\href{https://doi.org/10.18653/v1/2024.acl-long.818}{doi:\nolinkurl{10.18653/v1/2024.acl-long.818}}


\bibitem[Li et~al\mbox{.}(2024)]%
        {li2024distflashattndistributedmemoryefficientattention}
\bibfield{author}{\bibinfo{person}{Dacheng Li}, \bibinfo{person}{Rulin Shao}, \bibinfo{person}{Anze Xie}, \bibinfo{person}{Eric~P. Xing}, \bibinfo{person}{Xuezhe Ma}, \bibinfo{person}{Ion Stoica}, \bibinfo{person}{Joseph~E. Gonzalez}, {and} \bibinfo{person}{Hao Zhang}.} \bibinfo{year}{2024}\natexlab{}.
\newblock \bibinfo{title}{DISTFLASHATTN: Distributed Memory-efficient Attention for Long-context LLMs Training}.
\newblock
\showeprint[arxiv]{2310.03294}~[cs.LG]
\urldef\tempurl%
\url{https://arxiv.org/abs/2310.03294}
\showURL{%
\tempurl}


\bibitem[Liu et~al\mbox{.}(2023)]%
        {liu2023ringattentionblockwisetransformers}
\bibfield{author}{\bibinfo{person}{Hao Liu}, \bibinfo{person}{Matei Zaharia}, {and} \bibinfo{person}{Pieter Abbeel}.} \bibinfo{year}{2023}\natexlab{}.
\newblock \bibinfo{title}{Ring Attention with Blockwise Transformers for Near-Infinite Context}.
\newblock
\showeprint[arxiv]{2310.01889}~[cs.CL]
\urldef\tempurl%
\url{https://arxiv.org/abs/2310.01889}
\showURL{%
\tempurl}


\bibitem[Meta(2023)]%
        {touvron2023llama2openfoundation}
\bibfield{author}{\bibinfo{person}{Meta}.} \bibinfo{year}{2023}\natexlab{}.
\newblock \bibinfo{title}{Llama 2: Open Foundation and Fine-Tuned Chat Models}.
\newblock
\showeprint[arxiv]{2307.09288}~[cs.CL]
\urldef\tempurl%
\url{https://arxiv.org/abs/2307.09288}
\showURL{%
\tempurl}


\bibitem[Meta(2024)]%
        {grattafiori2024llama3herdmodels}
\bibfield{author}{\bibinfo{person}{Meta}.} \bibinfo{year}{2024}\natexlab{}.
\newblock \bibinfo{title}{The Llama 3 Herd of Models}.
\newblock
\showeprint[arxiv]{2407.21783}~[cs.AI]
\urldef\tempurl%
\url{https://arxiv.org/abs/2407.21783}
\showURL{%
\tempurl}


\bibitem[Narayanan et~al\mbox{.}(2021)]%
        {megatron2}
\bibfield{author}{\bibinfo{person}{Deepak Narayanan}, \bibinfo{person}{Mohammad Shoeybi}, \bibinfo{person}{Jared Casper}, \bibinfo{person}{Patrick LeGresley}, \bibinfo{person}{Mostofa Patwary}, \bibinfo{person}{Vijay Korthikanti}, \bibinfo{person}{Dmitri Vainbrand}, \bibinfo{person}{Prethvi Kashinkunti}, \bibinfo{person}{Julie Bernauer}, \bibinfo{person}{Bryan Catanzaro}, \bibinfo{person}{Amar Phanishayee}, {and} \bibinfo{person}{Matei Zaharia}.} \bibinfo{year}{2021}\natexlab{}.
\newblock \showarticletitle{Efficient large-scale language model training on GPU clusters using megatron-LM}. In \bibinfo{booktitle}{\emph{Proceedings of the International Conference for High Performance Computing, Networking, Storage and Analysis}} (St. Louis, Missouri) \emph{(\bibinfo{series}{SC '21})}. \bibinfo{publisher}{Association for Computing Machinery}, \bibinfo{address}{New York, NY, USA}, Article \bibinfo{articleno}{58}, \bibinfo{numpages}{15}~pages.
\newblock
\showISBNx{9781450384421}
\href{https://doi.org/10.1145/3458817.3476209}{doi:\nolinkurl{10.1145/3458817.3476209}}


\bibitem[NVIDIA(2022)]%
        {NCCL}
\bibfield{author}{\bibinfo{person}{NVIDIA}.} \bibinfo{year}{2022}\natexlab{}.
\newblock \bibinfo{title}{{NVIDIA Collective Communication Library (NCCL) Documentation}}.
\newblock \bibinfo{howpublished}{\url{https://docs.nvidia.com/deeplearning/nccl/user-guide/docs/index.html}}.
\newblock


\bibitem[NVIDIA(2025a)]%
        {a100}
\bibfield{author}{\bibinfo{person}{NVIDIA}.} \bibinfo{year}{2025}\natexlab{a}.
\newblock \bibinfo{title}{DGX A100 System Topology}.
\newblock
\urldef\tempurl%
\url{https://docs.nvidia.com/dgx/dgxa100-user-guide/introduction-to-dgxa100.html}
\showURL{%
\tempurl}


\bibitem[NVIDIA(2025b)]%
        {h100}
\bibfield{author}{\bibinfo{person}{NVIDIA}.} \bibinfo{year}{2025}\natexlab{b}.
\newblock \bibinfo{title}{DGX H100/200 System Topology}.
\newblock
\urldef\tempurl%
\url{https://docs.nvidia.com/dgx/dgxh100-user-guide/}
\showURL{%
\tempurl}


\bibitem[NVIDIA(2025c)]%
        {te}
\bibfield{author}{\bibinfo{person}{NVIDIA}.} \bibinfo{year}{2025}\natexlab{c}.
\newblock \bibinfo{title}{TransformerEngine}.
\newblock
\urldef\tempurl%
\url{https://github.com/NVIDIA/TransformerEngine}
\showURL{%
\tempurl}


\bibitem[Paster et~al\mbox{.}(2023)]%
        {paster2023openwebmathopendatasethighquality}
\bibfield{author}{\bibinfo{person}{Keiran Paster}, \bibinfo{person}{Marco~Dos Santos}, \bibinfo{person}{Zhangir Azerbayev}, {and} \bibinfo{person}{Jimmy Ba}.} \bibinfo{year}{2023}\natexlab{}.
\newblock \bibinfo{title}{OpenWebMath: An Open Dataset of High-Quality Mathematical Web Text}.
\newblock
\showeprint[arxiv]{2310.06786}~[cs.AI]
\urldef\tempurl%
\url{https://arxiv.org/abs/2310.06786}
\showURL{%
\tempurl}


\bibitem[Penedo et~al\mbox{.}(2024)]%
        {penedo2024finewebdatasetsdecantingweb}
\bibfield{author}{\bibinfo{person}{Guilherme Penedo}, \bibinfo{person}{Hynek Kydlíček}, \bibinfo{person}{Loubna~Ben allal}, \bibinfo{person}{Anton Lozhkov}, \bibinfo{person}{Margaret Mitchell}, \bibinfo{person}{Colin Raffel}, \bibinfo{person}{Leandro~Von Werra}, {and} \bibinfo{person}{Thomas Wolf}.} \bibinfo{year}{2024}\natexlab{}.
\newblock \bibinfo{title}{The FineWeb Datasets: Decanting the Web for the Finest Text Data at Scale}.
\newblock
\showeprint[arxiv]{2406.17557}~[cs.CL]
\urldef\tempurl%
\url{https://arxiv.org/abs/2406.17557}
\showURL{%
\tempurl}


\bibitem[Pouransari et~al\mbox{.}(2024)]%
        {applebucket}
\bibfield{author}{\bibinfo{person}{Hadi Pouransari}, \bibinfo{person}{Chun-Liang Li}, \bibinfo{person}{Jen-Hao~Rick Chang}, \bibinfo{person}{Pavan Kumar~Anasosalu Vasu}, \bibinfo{person}{Cem Koc}, \bibinfo{person}{Vaishaal Shankar}, {and} \bibinfo{person}{Oncel Tuzel}.} \bibinfo{year}{2024}\natexlab{}.
\newblock \showarticletitle{Dataset Decomposition: Faster LLM Training with Variable Sequence Length Curriculum}. In \bibinfo{booktitle}{\emph{Advances in Neural Information Processing Systems}}, \bibfield{editor}{\bibinfo{person}{A.~Globerson}, \bibinfo{person}{L.~Mackey}, \bibinfo{person}{D.~Belgrave}, \bibinfo{person}{A.~Fan}, \bibinfo{person}{U.~Paquet}, \bibinfo{person}{J.~Tomczak}, {and} \bibinfo{person}{C.~Zhang}} (Eds.), Vol.~\bibinfo{volume}{37}. \bibinfo{publisher}{Curran Associates, Inc.}, \bibinfo{pages}{36121--36147}.
\newblock
\urldef\tempurl%
\url{https://proceedings.neurips.cc/paper_files/paper/2024/file/3f9bf45ea04c98ad7cb857f951f499e2-Paper-Conference.pdf}
\showURL{%
\tempurl}


\bibitem[Qi et~al\mbox{.}(2023)]%
        {qi2023zerobubblepipelineparallelism}
\bibfield{author}{\bibinfo{person}{Penghui Qi}, \bibinfo{person}{Xinyi Wan}, \bibinfo{person}{Guangxing Huang}, {and} \bibinfo{person}{Min Lin}.} \bibinfo{year}{2023}\natexlab{}.
\newblock \bibinfo{title}{Zero Bubble Pipeline Parallelism}.
\newblock
\showeprint[arxiv]{2401.10241}~[cs.DC]
\urldef\tempurl%
\url{https://arxiv.org/abs/2401.10241}
\showURL{%
\tempurl}


\bibitem[Rajbhandari et~al\mbox{.}(2020)]%
        {rajbhandari2020zeromemoryoptimizationstraining}
\bibfield{author}{\bibinfo{person}{Samyam Rajbhandari}, \bibinfo{person}{Jeff Rasley}, \bibinfo{person}{Olatunji Ruwase}, {and} \bibinfo{person}{Yuxiong He}.} \bibinfo{year}{2020}\natexlab{}.
\newblock \bibinfo{title}{ZeRO: Memory Optimizations Toward Training Trillion Parameter Models}.
\newblock
\showeprint[arxiv]{1910.02054}~[cs.LG]
\urldef\tempurl%
\url{https://arxiv.org/abs/1910.02054}
\showURL{%
\tempurl}


\bibitem[Rasley et~al\mbox{.}(2020)]%
        {deepspeed}
\bibfield{author}{\bibinfo{person}{Jeff Rasley}, \bibinfo{person}{Samyam Rajbhandari}, \bibinfo{person}{Olatunji Ruwase}, {and} \bibinfo{person}{Yuxiong He}.} \bibinfo{year}{2020}\natexlab{}.
\newblock \showarticletitle{DeepSpeed: System Optimizations Enable Training Deep Learning Models with Over 100 Billion Parameters}. In \bibinfo{booktitle}{\emph{Proceedings of the 26th ACM SIGKDD International Conference on Knowledge Discovery \& Data Mining}} (Virtual Event, CA, USA) \emph{(\bibinfo{series}{KDD '20})}. \bibinfo{publisher}{Association for Computing Machinery}, \bibinfo{address}{New York, NY, USA}, \bibinfo{pages}{3505–3506}.
\newblock
\showISBNx{9781450379984}
\href{https://doi.org/10.1145/3394486.3406703}{doi:\nolinkurl{10.1145/3394486.3406703}}


\bibitem[Shen et~al\mbox{.}(2024)]%
        {shen2024slimpajamadcunderstandingdatacombinations}
\bibfield{author}{\bibinfo{person}{Zhiqiang Shen}, \bibinfo{person}{Tianhua Tao}, \bibinfo{person}{Liqun Ma}, \bibinfo{person}{Willie Neiswanger}, \bibinfo{person}{Zhengzhong Liu}, \bibinfo{person}{Hongyi Wang}, \bibinfo{person}{Bowen Tan}, \bibinfo{person}{Joel Hestness}, \bibinfo{person}{Natalia Vassilieva}, \bibinfo{person}{Daria Soboleva}, {and} \bibinfo{person}{Eric Xing}.} \bibinfo{year}{2024}\natexlab{}.
\newblock \bibinfo{title}{SlimPajama-DC: Understanding Data Combinations for LLM Training}.
\newblock
\showeprint[arxiv]{2309.10818}~[cs.CL]
\urldef\tempurl%
\url{https://arxiv.org/abs/2309.10818}
\showURL{%
\tempurl}


\bibitem[Shoeybi et~al\mbox{.}(2019)]%
        {megatron}
\bibfield{author}{\bibinfo{person}{Mohammad Shoeybi}, \bibinfo{person}{Mostofa Patwary}, \bibinfo{person}{Raul Puri}, \bibinfo{person}{Patrick LeGresley}, \bibinfo{person}{Jared Casper}, {and} \bibinfo{person}{Bryan Catanzaro}.} \bibinfo{year}{2019}\natexlab{}.
\newblock \showarticletitle{Megatron-LM: Training Multi-Billion Parameter Language Models Using Model Parallelism}.
\newblock \bibinfo{journal}{\emph{CoRR}}  \bibinfo{volume}{abs/1909.08053} (\bibinfo{year}{2019}).
\newblock
\showeprint[arXiv]{1909.08053}
\urldef\tempurl%
\url{http://arxiv.org/abs/1909.08053}
\showURL{%
\tempurl}


\bibitem[Smith et~al\mbox{.}(2018)]%
        {largebatch}
\bibfield{author}{\bibinfo{person}{Samuel~L. Smith}, \bibinfo{person}{Pieter-Jan Kindermans}, \bibinfo{person}{Chris Ying}, {and} \bibinfo{person}{Quoc~V. Le}.} \bibinfo{year}{2018}\natexlab{}.
\newblock \bibinfo{title}{Don't Decay the Learning Rate, Increase the Batch Size}.
\newblock
\showeprint[arxiv]{1711.00489}~[cs.LG]
\urldef\tempurl%
\url{https://arxiv.org/abs/1711.00489}
\showURL{%
\tempurl}


\bibitem[Team(2024)]%
        {yang2024qwen2technicalreport}
\bibfield{author}{\bibinfo{person}{Qwen Team}.} \bibinfo{year}{2024}\natexlab{}.
\newblock \bibinfo{title}{Qwen2 Technical Report}.
\newblock
\showeprint[arxiv]{2407.10671}~[cs.CL]
\urldef\tempurl%
\url{https://arxiv.org/abs/2407.10671}
\showURL{%
\tempurl}


\bibitem[Team(2025)]%
        {qwen2025qwen25technicalreport}
\bibfield{author}{\bibinfo{person}{Qwen Team}.} \bibinfo{year}{2025}\natexlab{}.
\newblock \bibinfo{title}{Qwen2.5 Technical Report}.
\newblock
\showeprint[arxiv]{2412.15115}~[cs.CL]
\urldef\tempurl%
\url{https://arxiv.org/abs/2412.15115}
\showURL{%
\tempurl}


\bibitem[Vaswani et~al\mbox{.}(2017)]%
        {attention}
\bibfield{author}{\bibinfo{person}{Ashish Vaswani}, \bibinfo{person}{Noam Shazeer}, \bibinfo{person}{Niki Parmar}, \bibinfo{person}{Jakob Uszkoreit}, \bibinfo{person}{Llion Jones}, \bibinfo{person}{Aidan~N Gomez}, \bibinfo{person}{\L~ukasz Kaiser}, {and} \bibinfo{person}{Illia Polosukhin}.} \bibinfo{year}{2017}\natexlab{}.
\newblock \showarticletitle{Attention is All you Need}. In \bibinfo{booktitle}{\emph{Advances in Neural Information Processing Systems}}, \bibfield{editor}{\bibinfo{person}{I.~Guyon}, \bibinfo{person}{U.~Von Luxburg}, \bibinfo{person}{S.~Bengio}, \bibinfo{person}{H.~Wallach}, \bibinfo{person}{R.~Fergus}, \bibinfo{person}{S.~Vishwanathan}, {and} \bibinfo{person}{R.~Garnett}} (Eds.), Vol.~\bibinfo{volume}{30}. \bibinfo{publisher}{Curran Associates, Inc.}
\newblock
\urldef\tempurl%
\url{https://proceedings.neurips.cc/paper_files/paper/2017/file/3f5ee243547dee91fbd053c1c4a845aa-Paper.pdf}
\showURL{%
\tempurl}


\bibitem[Wang et~al\mbox{.}(2025b)]%
        {flexsp}
\bibfield{author}{\bibinfo{person}{Yujie Wang}, \bibinfo{person}{Shiju Wang}, \bibinfo{person}{Shenhan Zhu}, \bibinfo{person}{Fangcheng Fu}, \bibinfo{person}{Xinyi Liu}, \bibinfo{person}{Xuefeng Xiao}, \bibinfo{person}{Huixia Li}, \bibinfo{person}{Jiashi Li}, \bibinfo{person}{Faming Wu}, {and} \bibinfo{person}{Bin Cui}.} \bibinfo{year}{2025}\natexlab{b}.
\newblock \showarticletitle{FlexSP: Accelerating Large Language Model Training via Flexible Sequence Parallelism}. In \bibinfo{booktitle}{\emph{Proceedings of the 30th ACM International Conference on Architectural Support for Programming Languages and Operating Systems, Volume 2}} (Rotterdam, Netherlands) \emph{(\bibinfo{series}{ASPLOS '25})}. \bibinfo{publisher}{Association for Computing Machinery}, \bibinfo{address}{New York, NY, USA}, \bibinfo{pages}{421–436}.
\newblock
\showISBNx{9798400710797}
\href{https://doi.org/10.1145/3676641.3715998}{doi:\nolinkurl{10.1145/3676641.3715998}}


\bibitem[Wang et~al\mbox{.}(2025a)]%
        {wlb}
\bibfield{author}{\bibinfo{person}{Zheng Wang}, \bibinfo{person}{Anna Cai}, \bibinfo{person}{Xinfeng Xie}, \bibinfo{person}{Zaifeng Pan}, \bibinfo{person}{Yue Guan}, \bibinfo{person}{Weiwei Chu}, \bibinfo{person}{Jie Wang}, \bibinfo{person}{Shikai Li}, \bibinfo{person}{Jianyu Huang}, \bibinfo{person}{Chris Cai}, \bibinfo{person}{Yuchen Hao}, {and} \bibinfo{person}{Yufei Ding}.} \bibinfo{year}{2025}\natexlab{a}.
\newblock \bibinfo{title}{WLB-LLM: Workload-Balanced 4D Parallelism for Large Language Model Training}.
\newblock
\showeprint[arxiv]{2503.17924}~[cs.DC]
\urldef\tempurl%
\url{https://arxiv.org/abs/2503.17924}
\showURL{%
\tempurl}


\bibitem[Xiong et~al\mbox{.}(2024)]%
        {llama2_mixtrue}
\bibfield{author}{\bibinfo{person}{Wenhan Xiong}, \bibinfo{person}{Jingyu Liu}, \bibinfo{person}{Igor Molybog}, \bibinfo{person}{Hejia Zhang}, \bibinfo{person}{Prajjwal Bhargava}, \bibinfo{person}{Rui Hou}, \bibinfo{person}{Louis Martin}, \bibinfo{person}{Rashi Rungta}, \bibinfo{person}{Karthik~Abinav Sankararaman}, \bibinfo{person}{Barlas Oguz}, \bibinfo{person}{Madian Khabsa}, \bibinfo{person}{Han Fang}, \bibinfo{person}{Yashar Mehdad}, \bibinfo{person}{Sharan Narang}, \bibinfo{person}{Kshitiz Malik}, \bibinfo{person}{Angela Fan}, \bibinfo{person}{Shruti Bhosale}, \bibinfo{person}{Sergey Edunov}, \bibinfo{person}{Mike Lewis}, \bibinfo{person}{Sinong Wang}, {and} \bibinfo{person}{Hao Ma}.} \bibinfo{year}{2024}\natexlab{}.
\newblock \showarticletitle{Effective Long-Context Scaling of Foundation Models}. In \bibinfo{booktitle}{\emph{Proceedings of the 2024 Conference of the North American Chapter of the Association for Computational Linguistics: Human Language Technologies (Volume 1: Long Papers)}}, \bibfield{editor}{\bibinfo{person}{Kevin Duh}, \bibinfo{person}{Helena Gomez}, {and} \bibinfo{person}{Steven Bethard}} (Eds.). \bibinfo{publisher}{Association for Computational Linguistics}, \bibinfo{address}{Mexico City, Mexico}, \bibinfo{pages}{4643--4663}.
\newblock
\href{https://doi.org/10.18653/v1/2024.naacl-long.260}{doi:\nolinkurl{10.18653/v1/2024.naacl-long.260}}


\bibitem[You et~al\mbox{.}(2017)]%
        {lars}
\bibfield{author}{\bibinfo{person}{Yang You}, \bibinfo{person}{Igor Gitman}, {and} \bibinfo{person}{Boris Ginsburg}.} \bibinfo{year}{2017}\natexlab{}.
\newblock \bibinfo{title}{Large Batch Training of Convolutional Networks}.
\newblock
\showeprint[arxiv]{1708.03888}~[cs.CV]
\urldef\tempurl%
\url{https://arxiv.org/abs/1708.03888}
\showURL{%
\tempurl}


\bibitem[You et~al\mbox{.}(2020)]%
        {lamb}
\bibfield{author}{\bibinfo{person}{Yang You}, \bibinfo{person}{Jing Li}, \bibinfo{person}{Sashank Reddi}, \bibinfo{person}{Jonathan Hseu}, \bibinfo{person}{Sanjiv Kumar}, \bibinfo{person}{Srinadh Bhojanapalli}, \bibinfo{person}{Xiaodan Song}, \bibinfo{person}{James Demmel}, \bibinfo{person}{Kurt Keutzer}, {and} \bibinfo{person}{Cho-Jui Hsieh}.} \bibinfo{year}{2020}\natexlab{}.
\newblock \bibinfo{title}{Large Batch Optimization for Deep Learning: Training BERT in 76 minutes}.
\newblock
\showeprint[arxiv]{1904.00962}~[cs.LG]
\urldef\tempurl%
\url{https://arxiv.org/abs/1904.00962}
\showURL{%
\tempurl}


\bibitem[Yuan et~al\mbox{.}(2024)]%
        {kuaishouatc24}
\bibfield{author}{\bibinfo{person}{Tailing Yuan}, \bibinfo{person}{Yuliang Liu}, \bibinfo{person}{Xucheng Ye}, \bibinfo{person}{Shenglong Zhang}, \bibinfo{person}{Jianchao Tan}, \bibinfo{person}{Bin Chen}, \bibinfo{person}{Chengru Song}, {and} \bibinfo{person}{Di Zhang}.} \bibinfo{year}{2024}\natexlab{}.
\newblock \showarticletitle{Accelerating the Training of Large Language Models using Efficient Activation Rematerialization and Optimal Hybrid Parallelism}. In \bibinfo{booktitle}{\emph{2024 USENIX Annual Technical Conference (USENIX ATC 24)}}. \bibinfo{publisher}{USENIX Association}, \bibinfo{address}{Santa Clara, CA}, \bibinfo{pages}{545--561}.
\newblock
\showISBNx{978-1-939133-41-0}
\urldef\tempurl%
\url{https://www.usenix.org/conference/atc24/presentation/yuan}
\showURL{%
\tempurl}


\bibitem[Zhao et~al\mbox{.}(2024)]%
        {zhao2024chunk}
\bibfield{author}{\bibinfo{person}{Liang Zhao}, \bibinfo{person}{Tianwen Wei}, \bibinfo{person}{Liang Zeng}, \bibinfo{person}{Cheng Cheng}, \bibinfo{person}{Liu Yang}, \bibinfo{person}{Peng Cheng}, \bibinfo{person}{Lijie Wang}, \bibinfo{person}{Chenxia Li}, \bibinfo{person}{Xuejie Wu}, \bibinfo{person}{Bo Zhu}, \bibinfo{person}{Yimeng Gan}, \bibinfo{person}{Rui Hu}, \bibinfo{person}{Shuicheng Yan}, \bibinfo{person}{Han Fang}, {and} \bibinfo{person}{Yahui Zhou}.} \bibinfo{year}{2024}\natexlab{}.
\newblock \bibinfo{title}{LongSkywork: A Training Recipe for Efficiently Extending Context Length in Large Language Models}.
\newblock
\showeprint[arxiv]{2406.00605}~[cs.CL]
\urldef\tempurl%
\url{https://arxiv.org/abs/2406.00605}
\showURL{%
\tempurl}


\end{thebibliography}
\end{document}